\documentclass[sn-basic]{sn-jnl}

\usepackage{graphicx}%
\usepackage{multirow}%
\usepackage{amsmath,amssymb,amsfonts}%
\usepackage{amsthm}%
\usepackage{mathrsfs}%
\usepackage[title]{appendix}%
\usepackage{xcolor}%
\usepackage{textcomp}%
\usepackage{manyfoot}%
\usepackage{booktabs}%
\usepackage{algorithm}%
\usepackage{algorithmicx}%
\usepackage{algpseudocode}%
\usepackage{listings}%
\usepackage{natbib}

\graphicspath{{graficos}}
\begin{document}

\title{Spatiotemporal Risk-Based Hotspot Dynamics Assessment}

\author[1]{\fnm{Jos\'e M.} \sur{Angulo}}\email{jmangulo@ugr.es}
\equalcont{These authors contributed equally to this work.}

\author*[1]{\fnm{Ana E.} \sur{Madrid}}\email{anaesther@ugr.es}
\equalcont{These authors contributed equally to this work.}

\affil*[1]{\orgdiv{Department of Statistics and Operations Research}, \orgname{University of Granada}, \orgaddress{\street{Campus Fuente Nueva s/n}, \city{Granada}, \postcode{18071}, \country{Spain}}}

\abstract{

Analysis of hotspots often constitutes the final objective in application studies oriented to spatiotemporal risk assessment, particularly for risk mapping and as a helpful instrument for decision making, in diverse fields such as Ecology, Epidemiology, Environmental Sciences, Security, among others. The most primary form, common in practice, consists in identifying and depicting the areas where the magnitude of interest exceeds a certain critical threshold level. In this work, we introduce a wider, more general concept of `hotspot', 
associated with the extremal values of risk measures defined on different structural indicators of threshold exceedance sets, such as the exceedance area, the excess volume, etc. In particular, we propose a methodological approach based on empirical characterization of the indicator probability distributions from simulation conditional on the available observations, calculated at local resolution scales determined by sliding windows. Risk map generation is then supported by a proper conceptualization and formalization of a suitable family of risk conditional random fields (RCRFs), specified in terms of a set of input elements that can be viewed as  flexible tuning parameters for practical implementation and application.  
For illustration, we consider an extension of a spatiotemporal diffusive blur-generated autoregressive model involving a dynamical spatial deformation. Evolutionary assessment of hotspots is then addressed under different scenarios, with reference to well-known quantile-based risk measures. 
}


\maketitle

\section{Introduction}
\label{intro}

Identification and qualitative-quantitative statistical characterization of
hotspots, representing clustered areas where a magnitude of interest subject to
spatiotemporal dynamics exceeds prescribed critical levels, constitutes a
prominent objective in many studies oriented to risk assessment in diverse areas
of application, such as Ecology and Environmental Sciences, Epidemiology,
Security, etc. Different adverse phenomena with impact on systems and
populations are, in fact, related to extremal occurrences in meteorological,
climatic, or even anthropogenic variables (e.g., \citeauthor{Lawson2010} \citeyear{Lawson2010};
\citeauthor{Leibovici2011} \citeyear{Leibovici2011}; \citeauthor{Butt2020} \citeyear{Butt2020};
\citeauthor{VieiraPassos2024} \citeyear{VieiraPassos2024}). A referential prospective
motivation for this analytical approach was introduced in
\citeauthor{Patil2010a} (\citeyear{Patil2010a}, \citeyear{Patil2010b}).

The classical approach to hotspot dynamics assessment is directly focused on the analysis and graphical representation of exceedances of the spatiotemporal magnitude of interest above a given threshold. That is, in this setting, hotspots are identified with the connected components of threshold exceedance sets, which are jointly visualized and subsequently interpreted in the form of risk maps as an output for practical evaluation and potential decision making.     

However, in many cases, it will be interesting to describe the extremal behavior occurrences in terms of diverse, more general indicators; among them, indicators based on topological and geometrical characteristics of excursion sets, such as intrinsic volumes (or, related to them, Lipschitz-Killing curvatures, Minkowski functionals, etc.), have received special attention in the literature; see, e.g., \citet{AdlerTaylor2007,AdlerTaylor2011}, \citet{AzaisWschebor2009} and \citet{Vanmarcke2010}. Some examples of practical interest include, under  different versions, fragmentation in terms of connected components, perimetral lengths (see, e.g., \citeauthor{Cotsakis2024} \citeyear{Cotsakis2024}), exceedance areas or excess volumes (see, e.g., 
\citeauthor{AnguloMadrid2010} \citeyear{AnguloMadrid2010}, \citeyear{AnguloMadrid2014}, \citeyear{AnguloMadrid2023};
\citeauthor{MadridAnguloMateu2012} \citeyear{MadridAnguloMateu2012}, \citeyear{MadridAnguloMateu2016};
\citeauthor{RomeroMadridAngulo2018} \citeyear{RomeroMadridAngulo2018}), as well as higher-order indicators (see, e.g., \citeauthor{Christakos1996} \citeyear{Christakos1996}, \citeyear{Christakos1997}).             

Given the random nature of excursion sets, and hence, of the different characterizing structural indicators, under the idea that in many cases such indicators can be naturally interpreted as magnitudes intrinsically representing a negative effect with respect to the phenomenon considered, these can be objectively adopted as loss functions, that is, as the formal argument of different risk measures. In this context, in recent decades, there has been an intensive development of an advanced theory, with solid foundations, of risk measures (see, for example, 
\citeauthor{KluppelbergStraubWelpe2014} \citeyear{KluppelbergStraubWelpe2014}; \citeauthor{FollmerSchied2016} \citeyear{FollmerSchied2016}). In particular, among other aspects, suitable axiomatics have been established as a conceptual and formal basis for the definition of certain classes of risk measures, e.g., coherent (\citeauthor{Artzner1999} \citeyear{Artzner1999}) and convex (\citeauthor{FollmerSchied2002} \citeyear{FollmerSchied2002};
\citeauthor{FrittelliGianin2002} \citeyear{FrittelliGianin2002})  risk measures, and diverse families of risk measures have been introduced. Among the latter, notably stand out, due to their intuitive interpretation, quantile-based risk measures, such as Value-at-Risk (VaR) or Average Value-at-Risk (AVaR), and more generally spectral risk measures (see, e.g., \citeauthor{BelliniKlarMuellerRosazzaGianin2014} \citeyear{BelliniKlarMuellerRosazzaGianin2014}; \citeauthor{FollmerSchied2016} \citeyear{FollmerSchied2016}).          

In this work, we propose a flexible methodology for the analysis of 
\emph{risk-related hotspots}, here understood as the exceedances, over specified critical levels, of risk measures evaluated on selected indicators associated with excursion sets. Analysis and quantification of such indicators is performed at subregional scales defined in terms of sliding windows. 
The implementation of this methodology requires, as a preliminary step, the availability of: (i) a suitable model representing the phenomenon under study, either from prior knowledge or being previously estimated; and (ii) a set of observations, conditional on which simulated realizations can be derived from such a model. Intensive simulation allows then the empirical approximation of the conditional distributions of selected threshold-exceedance indicators at subregional scale, noting that, possibly except for very specific cases, these distributions will be analytically unknown. 
The empirical conditional distributions thus obtained constitute the basis for evaluation of risk measures, leading to a conceptual and formal definition of an underlying related family of \emph{risk conditional random fields} (RCRFs), whose realizations provide the generation of risk maps, with different inherent interpretations according to the mentioned input specifications. In the spatiotemporal case, this approach confers a character of \emph{predictive-risk assessment} at prescribed time horizons, eventually enabling adaptive implementation and update-output derivations based on real-time incoming observations.              

To illustrate the practical implementation and elements involved in the methodology proposed, and subsequent interpretations, we take for reference the extended version of \citet{Brown2000} blur-generated autoregressive spatiotemporal model introduced in \citet{AnguloMadrid2014} by integration of the dynamical effect of spatial deformation. This model is particularly interesting for our purpose since it allows incorporating local dynamical heterogeneity, visualizing its effect in terms of the dynamics of predictive risk-related hotspots, for different scenarios in reference to the underlying process and the input specifications involved in the risk conditional random field.       

The remainder of the paper is structured as follows. 
Section~\ref{sec:preliminaries} introduces the preliminary concepts and notation 
required for the subsequent developments, including geometrical-probabilistic 
indicators associated with threshold exceedance sets, basic notions on risk measures, 
and aspects related to spatial deformation of random fields. 
Section~\ref{sec:methodology} presents in detail the proposed methodological framework 
for spatiotemporal risk-based hotspot assessment, describing its conceptual basis and the main steps involved in its implementation.
In Section~\ref{sec:Illustration}, the methodology is illustrated through a simulation 
study under different scenarios, highlighting its flexibility and interpretative 
capabilities. 
Finally, Section~\ref{sec:conclusions} summarizes the main findings of the paper and 
discusses some perspectives and open directions for future research.

\section{Preliminaries}\label{sec:preliminaries}

In this section, we introduce preliminary contents related to geometrical-probabilistic indicators representing characteristics of threshold random field exceedance sets, risk measures, in particular, quantile-based formulations, and spatially deformed random fields. These aspects are fundamental  regarding the methodology proposed and illustrations derived in this paper.

\subsection{Random field threshold exceedances}\label{subsec:indicators}

Let $X=\{X(\mathbf{s}): \mathbf{s}\in S\subseteq\mathbb{R}^d\}$ be a spatial random field, and let $D\subseteq S$ be a bounded subdomain with non-null Lebesgue measure, i.e., $\lambda(D)>0$. For simplicity, we assume that all random variables $X(\mathbf{s})$ have continuous distributions. The random \textit{excursion set} of $X$ in $D$ over the constant threshold $u\in\mathbb{R}$ is defined as 
$$A_u(X,D)=\{\mathbf{s}\in D: X(\mathbf{s})\geq u\}.$$
A comprehensive exploration of the geometry of random field excursion sets, focusing on intrinsic volumes, is provided by \citeauthor{AdlerTaylor2007} (\citeyear{AdlerTaylor2007}, \citeyear{AdlerTaylor2011}); see also \citet{AzaisWschebor2009} and \citet{Vanmarcke2010}.

In this work, we consider the following indicators related to exceedance areas and excess volumes, assuming that  $D$ is compact and the random field $X$ satisfies suitable regularity and measurability conditions:
\begin{itemize}
\item The \textit{relative exceedance area with respect to the subdomain D},
$$ \lambda^D(A_u(X,D))=\frac{\lambda(A_u(X,D))}{\lambda(D)},$$
where $\lambda(A_u(X,D))=\int_{A_u(X,D)}d\mathbf{s}$ represents the total area where the random field values exceed the threshold $u$.

\item The \textit{relative excess volume with respect to the area of the subdomain $D$}, 
$$V^D(A_u(X,D))=\frac{V(A_u(X,D))}{\lambda(D)},$$
where $V(A_u(X,D))=\int_u^{+\infty}{ \lambda (A_{u^{\prime}}(X,D))}du^{\prime}$ denotes the total excess volume of the random field above the threshold $u$.
\end{itemize}

The ratio of these two indicators defines the \textit{relative excess volume with respect to the area of the excursion set},
$$V^{A_u}(A_u(X,D))=\frac{V(A_u(X,D))}{\lambda(A_u(X,D))}=\frac{V^D(A_u(X,D))}{\lambda^D(A_u(X,D))}.$$

Expectations of the above `first-order' (i.e., single-location based) indicators can be suitably expressed in terms of the \textit{compound} cumulative distribution function (CDF).   

Let $F_X^D$ be the compound CDF of random field $X$ on a bounded subset $D\subseteq S$, defined as
$$F_X^D(x)=\int_D F_{X(\mathbf{s})}(x)\lambda^D(d\mathbf{s}), \quad \forall x \in \mathbb{R},$$     
where $F_{X(\mathbf{s})}$ denotes the CDF of $X(\mathbf{s})$, and $\lambda^D(\cdot):=\frac{\lambda(\cdot)}{\lambda(D)}$ represents the normalized Lebesgue measure on $D$. Let $S_X^D(x)=1-F_X^D(x)$, $x\in \mathbb{R}$, be the corresponding compound decumulative distribution function (DDF). Furthermore, let $X^D$ denote the compound random variable associated with $F_X^D$. The expected values of the relative exceedance area and the relative excess volume of the random field $X$ over the threshold $u$ in the subdomain $D$ can be expressed as
\begin{equation} \label{eq:expected_area}
E[\lambda^D(A_u(X,D))]=1-F_X^D(u)=S_X^D(u)
\end{equation}
and 
$$E[V^D(A_u(X,D))]=\int_u^\infty S_X^D (u')du'=E[X_u^D],$$
where $X_u^D=\max\{0, X^D-u\}$. 

Further details can be found in \citet{RomeroMadridAngulo2018}. 

\subsection{Risk measures}\label{subsec:risk-measures}
Given a probability space $(\Omega,\mathcal{F},P)$,  a \textit{risk measure} $\rho$ is a functional defined as
$$ \rho: \mathcal{M}(\Omega,\mathcal{F},P)\longrightarrow\bar{\mathbb{R}}$$
which satisfies certain suitable conditions. Here, $\mathcal{M}(\Omega,\mathcal{F},P)$ is a linear subspace of $\mathcal{M_B}(\Omega,\mathcal{F},P)$, the space of real-valued Borel-measurable functions defined on $(\Omega,\mathcal{F},P)$, and its elements are interpreted as `loss' functions. The notation $\bar{\mathbb{R}}$ indicates the compactification of $\mathbb{R}$ defined as $\mathbb{R}\cup\{-\infty,+\infty\}$. It is common for $\mathcal{M}(\Omega,\mathcal{F},P)$ to be considered as an $L^p(\Omega,\mathcal{F},P)$ space with $p\in [1,\infty]$.

Risk measures based on moments or quantiles are commonly used due to their direct interpretation regarding the distribution of the argument random variable. In this study, we focus on the well-known quantile-based risk measures named as \emph{Value-at-Risk} (VaR) and \emph{Average Value-at-Risk} (AVaR), defined below.

Let $L$ be a loss random variable and $\alpha \in (0,1]$. The Value-at-Risk at confidence level $1-\alpha$ of $L$, denoted $VaR_{1-\alpha}(L)$, is defined as the lower $(1-\alpha)-$quantile of the distribution of $L$:
\begin{equation*}
	VaR_{1-\alpha}(L)=\inf\{l\in \mathbb{R} \text{ : } F_{L}(l)\geq 1-\alpha\},  
\end{equation*} 
which is interpreted as the minimum loss that will occur in the $100\alpha\%$ less favorable scenarios.

The Average Value-at-Risk at confidence level $1-\alpha$ of $L$, denoted $AVaR_{1-\alpha}(L)$, is defined as
\begin{equation*}
	AVaR_{1-\alpha}(L) = \frac{1}{\alpha} \int_{1-\alpha}^1 VaR_{p}(L) dp.
\end{equation*}
If $L$ is a continuous random variable, it follows that $	VaR_{1-\alpha}(L)$ is equal to the infimum of the set $F_L^{-1}(1-\alpha)$. In this case, $AVaR_{1-\alpha}(L)$ is the conditional expectation
\begin{equation*}
	AVaR_{1-\alpha}(L) = E[L|L>VaR_{1-\alpha}(L)],
\end{equation*}
which represents the expected loss in the $100\alpha\%$ most adverse scenarios.

Some references on general aspects of risk measures are, for example, \citet{KrieleWolf2014}, \citeauthor{FollmerSchied2002} (\citeyear{FollmerSchied2002}, \citeyear{FollmerSchied2016}), and \citet{KluppelbergStraubWelpe2014}. Further exploration of risk measures based on  generalized quantiles can be found, for instance, in \citet{BelliniKlarMuellerRosazzaGianin2014}; see extremile-based alternatives in \citet{DaouiaGijbelsStupfler2019} and references therein.

\subsection{Deformation}\label{subsec:deformation}

Let $X$ be a random field on $S\subseteq\mathbb{R}^d$, and let $\Phi$ be a spatial deformation, defined as a one-to-one transformation $\Phi:S\longrightarrow S^{\prime}\subseteq\mathbb{R}^d$, such that $\Phi$ is a $C^1$-diffeomorphism with  strictly positive Jacobian. The `deformed' random field $X[\Phi] = \{X[\Phi](\mathbf{s}^{\prime}): \mathbf{s}^{\prime} \in S^{\prime} \}$ is given by
\begin{equation}\nonumber
X[\Phi](\mathbf{s}^{\prime})=X(\Phi^{-1}(\mathbf{s}^{\prime})),  \quad \forall \mathbf{s}^{\prime} \in S^{\prime}.
\end{equation}
This common interpretation of a random field deformation implies that the state values remain the same as in the original field, but are reallocated according to the deformation $\Phi$. The random field structure is then distorted due to the change in the geometry of the domain. In this context, the random field is assumed to represent a `level'-type physical magnitude. A formal definition of a transformed random field corresponding to a `flow'-type magnitude can be found, for example, in \citeauthor{AnguloMadrid2010} (\citeyear{AnguloMadrid2010}, \citeyear{AnguloMadrid2014}, \citeyear{AnguloMadrid2023}) and \citet{MadridAnguloMateu2016}. In a spatiotemporal framework, \citet{AnguloMadrid2014} consider the propagated effect of a dynamical spatial deformation.

\section{Methodology}\label{sec:methodology}

In this section, we describe in detail the methodological approach proposed in this work. 

Let $X(\mathbf{s},t)$ be a spatiotemporal random field on $S\times T\subseteq\mathbb{R}^d\times\mathbb{Z}$, representing the phenomenon under study. Let $\mathbf{X}_0$, on a certain subdomain $D\subseteq S$, denote the set of observations of $X$ available up to the present time, taken as $t=0$, and let us  generate a sufficiently large number $M$ of independent simulated realizations of the conditional process derived from $X(\mathbf{s},t)$,  based on the observed data, denoted as $X|\mathbf{X}_0$, over $N$ prediction time periods, $t=1,\ldots,N$. These realizations are represented as $(X|\mathbf{X}_0)^{[m]}$ for $m=1,\ldots , M$. The first step consists of the specification of the reference threshold level used for risk assessment over time.

\subsection{Threshold selection to define random field excursion sets}\label{subsec:threshold}

We will follow two possible strategies for determining the reference threshold defining the excursion set.

\medskip
 
\textbf{Method I}: In this approach, a fixed threshold is associated with the $(1-\alpha)$-quantile, where $\alpha \in (0,1)$, based on the observation $\mathbf{X}_0$. Specifically,
$$u_{1-\alpha}:= (\hat{F}_{\mathbf{X}_0}^D)^{-1}(1-\alpha)^- ,$$
where $ \hat{F}_{\mathbf{X}_0}^D$ is the empirical CDF from $\mathbf{X}_0$, and `$-$' stands for the infimum of the set.

\medskip

\textbf{Method II}: In this case, a time-varying threshold for each time point $t=1,\ldots,N$,  is associated with the $(1-\alpha)$-quantile, where $\alpha\in(0,1)$, based on $\mathbf{X}_t = \{(X(\cdot,t)|\mathbf{X}_0)^{[m]}: \ m=1,\ldots , M\}$. That is, at each time point, the threshold is defined by the corresponding $(1-\alpha)$-quantile of the empirical CDF $\hat{F}_{\mathbf{X}_t}^D$ obtained by aggregation of the $M$ conditional simulation samples at time $t$,
$$u_{t,1-\alpha} := (\hat{F}_{\mathbf{X}_t}^D)^{-1}(1-\alpha)^-.$$
In this way, from  equation (\ref{eq:expected_area}),  it follows that
$$E[\lambda^D(A_{u_{t,1-\alpha}}(\mathbf{X}_t,D))] \approx 1-\hat{F}_{\mathbf{X}_t}^D(u_{t,1-\alpha})=\alpha,\quad \forall t=1,\ldots,N$$
(see \citeauthor{RomeroMadridAngulo2018} \citeyear{RomeroMadridAngulo2018}).\\

Thus, in Method I, the threshold is fixed across all time points, making it appropriate when the spatial distribution of the variable of interest is relatively stable over time and when maintaining a consistent reference threshold is a priority. 
Under this approach, the proportion of the area in $D$ where the threshold $u_{1-\alpha}$ is exceeded is equal to $\alpha$ only for $t=0$. 
Conversely, Method II introduces a dynamic threshold that adjusts at each time point based on the distribution at that moment; that is, the threshold varies at each time point $t$, while the expected proportion of the area in $D$ where the threshold $u_{t,1-\alpha}$ is exceeded remains constant and equal to $\alpha$ at all times. This method may be particularly suited for contexts in which the underlying process displays significant temporal variations.\\

\subsection{Empirical evaluation of risk measures on exceedance indicators}\label{subsec:empiricalRM}

Let $\mathcal{I}_{1-\alpha}$ represent any first-order exceedance indicator such as those introduced in Subsection \ref{subsec:indicators}, formulated as a function of the realizations of the conditional spatiotemporal random field $X|\mathbf{X}_0$ on $D$, for each $t=1,\dots,N$. More precisely, these indicators are expressed as a function of the random excursion set family $A_{u_{1-\alpha'}}(X|\mathbf{X}_0,D)$ for $\alpha'\geq\alpha$, at each horizon time $t$. 
Now, for the simulated realizations $(X|\mathbf{X}_0)^{[m]}$, $m=1,\dots,M$, we can identify the corresponding excursion sets $\{A_{u_{1-\alpha'}}((X|\mathbf{X}_0)^{[m]},D): \ m=1,\ldots,M\}$ and the associated values of the indicator, $\mathcal{I}_{1-\alpha}^M = \{\mathcal{I}_{1-\alpha}^{[m]}: \ m=1,\ldots,M\}$.  Predictive risk is then assessed, for each time $t$, using the empirical CDF $\hat{F}_{\mathcal{I}_{1-\alpha}^M}$ derived from these realizations. 
For instance, if we define a threshold level as $u_{1-\alpha_1}$ for a specified exceedance area proportion $\alpha_1$, we can calculate the corresponding `empirical' VaR and AVaR values of the indicator $\mathcal{I}_{1-\alpha_1}$ at any confidence level $1-\alpha_2$, expressed as $VaR_{1-\alpha_2}( \mathcal{I}_{1-\alpha_1})$ and $AVaR_{1-\alpha_2}(\mathcal{I}_{1-\alpha_1})$.\\

\subsection{Local assessment in terms of a risk conditional RF}
\label{subsec:local-assessment}

For a fixed time $t$ and a threshold level $u$, we  conduct a local spatial analysis concerning the spatial distribution of the exceedances, following the previously outlined procedure applied to smaller subregions $D'$ of the entire domain. For simplicity, here we assume that $D$ is a square region and $D'$ is determined by a sliding square window of fixed size across $D$, with regular steps and partial overlapping along the coordinate axes. In practice, the computation of the selected indicators and risk measures, as described in Subsection \ref{subsec:empiricalRM}, is  performed for each subregion $D'$, enabling the construction of `local risk' maps on $D$ at various levels of resolution.\\

The elements and methodological aspects introduced above can be properly formalized based on the definition of a \emph{risk conditional random field} (RCRF). The source of randomness intrinsic to the latter is associated with the conditional simulation design; i.e., $M$ independent simulated realizations of $X|\mathbf{X}_0$, which are themselves random in nature, are used as the basis for risk assessment.

Since the final output pursued is a certain `risk map', this can be viewed as a realization of the RCRF 
$$RC[\Psi](\mathbf{c},t),$$
where $\mathbf{c}$ denotes the centroid of each window $D'$, and $t$ represents the predictive horizon. Here, $[\Psi]\equiv[\mathbf{X}_0,  M, \text{WD}, u, \mathcal{I}, \rho]$ stands for the set of specifications involved in the methodology, that is:
\begin{itemize}
    \item $\mathbf{X}_0$: the set of observed values of $X$ on a certain subdomain $D$ up to the present time ($t=0$).
    \item $M$: the number of independent conditional simulations generated from $X|\mathbf{X}_0$.
    \item WD: the sliding window design used for local assessment; in our case, in terms of size, step, and overlap.
    \item $u$: the reference threshold level used to define excursion sets.
    \item $\mathcal{I}$: the selected first-order exceedance indicator derived from the conditional realizations of the random field.
    \item $\rho$: the risk measure applied to the indicator, such as VaR or AVaR.
\end{itemize}

\subsection{Identification of risk-related hotspots}
\label{subsec:hotspots}
Once local risk maps are generated, based on a specified RCRF, hotspots can be defined as regions associated with the set of centroids where the values of $RC[\Psi](\mathbf{c},t)$ exceed a predetermined critical level $v$. Specifically, at each horizon time $t$, hotspots are identified with the class
\begin{equation*}
H_v(RC[\Psi], D, t) \equiv \pi_0\left(\left\{\mathbf{c} \in D: \/ RC[\Psi](\mathbf{c},t) \geq v  \right\}\right),
\end{equation*}
where $\pi_0(\cdot)$ denotes the class of connected components of a set.

It is important to emphasize that hotspots, under the approach proposed, are associated with excursion sets defined by the RCRF, rather than directly by the physical variable of interest. Consequently, the reference critical level $v$ can be determined using methodologies analogous to those described in Subsection \ref{subsec:threshold}. This procedure facilitates the identification of areas where extreme values indicate an elevated risk, enabling targeted interventions and management strategies. The critical level specification therefore induces the hotspots topological structure and dynamics.

\section{Illustration based on simulations}\label{sec:Illustration}

For illustration, we consider an extension of a spatiotemporal diffusive blur-generated autoregressive model involving a dynamical spatial deformation, introduced in \citet{AnguloMadrid2014}, given by

\begin{align}\label{model}
X(\mathbf{s},t)&=aX[\Phi;h](\mathbf{s},t-1)+Z(\mathbf{s},t)\nonumber \\
&=a\int_{\mathbb{R}^d}h(\mathbf{s}-\mathbf{s}')X(\Phi^{-1}(\mathbf{s}'),t-1)d\mathbf{s}'+Z(\mathbf{s},t),
\end{align}
with $X(\cdot,0)=X_0$ being a spatial random field on $\mathbb{R}^d$, $Z$ a spatiotemporal random field on $\mathbb{R}^d\times\mathbb{Z}$, $\Phi$ a spatial deformation, $h$ a blurring kernel, and $a$ a constant such that $0<a<1$. Specifically, here $Z$ is assumed to be uncorrelated in time, with spatial covariance given by the Cauchy model with $(\alpha,\beta)=(2,0.1)$, $\Phi$ is a deformation  with a dynamical effect reflected in a certain clockwise rotational movement, jointly with a progressive attraction towards the domain central point, $h$ is a homogeneous Gaussian kernel with variance $\sigma^2_h=10$, and $a=1-10^{-6}$.

This model has been selected due to its structural and dynamic characteristics. In particular, deformation induces a spatially non-homogeneous behavior, with a smooth change in temporal evolution preserving a certain stability from autoregression. This effect will be reflected in the dynamics of the RCRFs and related hotspots.   

The main purpose of this section is to describe in detail the elements and steps involved in the methodology proposed. Various scenarios are selected in particular to show the flexibility that this approach provides for practical applications, as well as the most significant related interpretations.  

In our case, given the nature of model (\ref{model}), we consider as the observation set $\mathbf{X}_0$ a complete realization at time $t=0$ on the square $[0,400]^2$, with each unit cell representing a pixel. This realization has been generated after running the dynamic mechanism during 150 times in order to achieve a stable regime removing the effect of initial conditions.  

Based on $\mathbf{X}_0$, $M=100$ realizations of the conditional random field $X|\mathbf{X}_0$ at horizon times $t=1, \dots, 10$ are simulated. For local assessment, we use a square sliding window of $33\times 33$ pixels, with window centers placed on a regular grid of step 11 pixels covering the whole domain (from pixel 1 to pixel 397 along each axis), and with windows truncated at the domain boundary. This results in a total of $37\times 37=1369$ tested subdomains, and determines the level of resolution adopted for risk mapping.

\begin{figure}[ht!]
    \centering
    \includegraphics[width=0.33\textwidth]{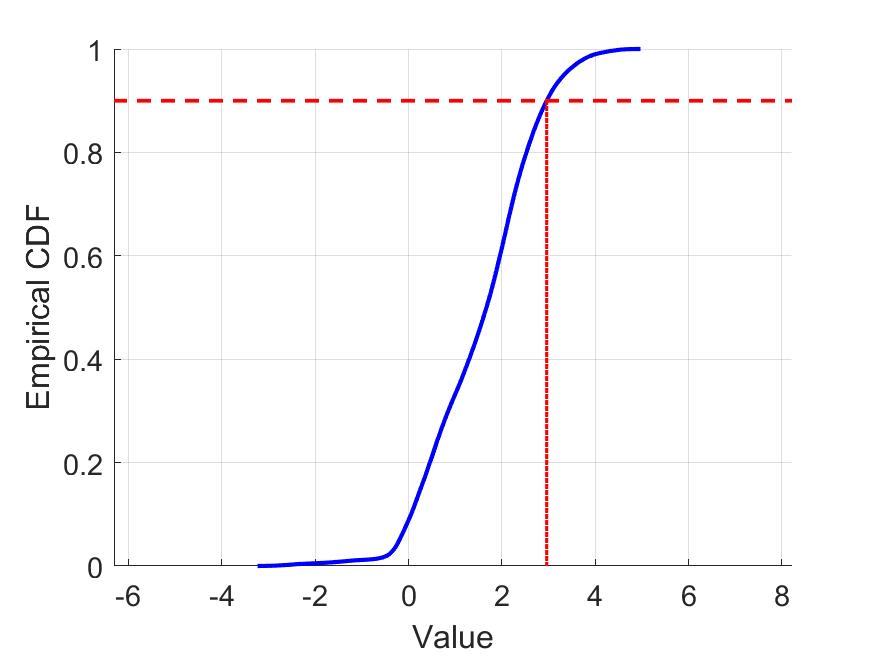}%
    \hskip 5mm
    \includegraphics[width=0.33\textwidth]{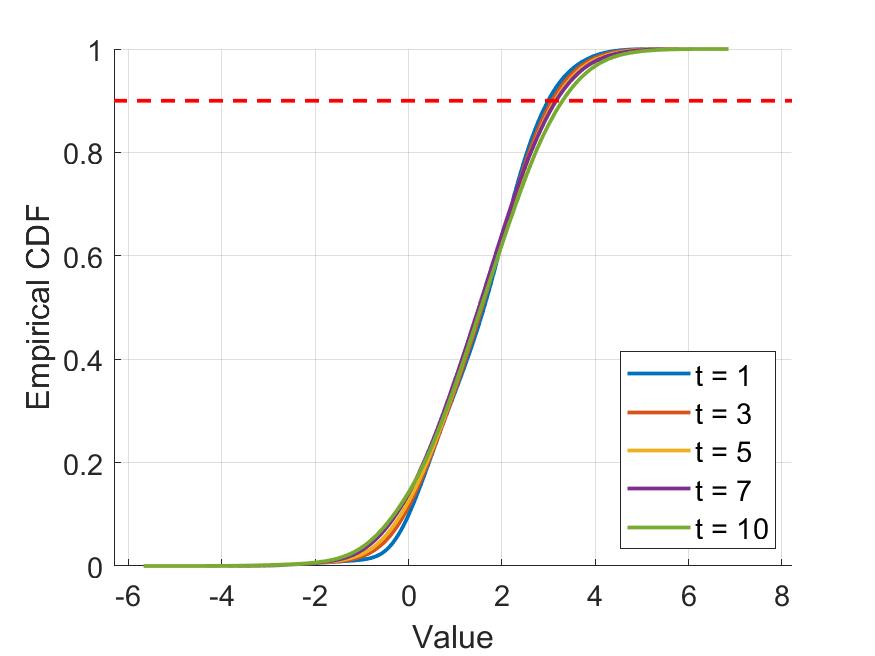}\\
  \caption{Empirical CDFs employed to determine the thresholds according to Methods I (left) and II (right), as detailed in Section~\ref{subsec:threshold}}
    \label{fig:threshold_selection}
\end{figure}

To specify the threshold level defining excursion sets, we inspect the empirical CDFs obtained from the observation set $\mathbf{X}_0$, under Method I (see left plot in Fig. \ref{fig:threshold_selection}), and the compound empirical CDFs obtained from the simulated realizations $\mathbf{X}_t$ at each horizon time $t$, under Method II (see right plot in Fig. \ref{fig:threshold_selection}, in particular, for times $t=1,3,5,7,10$). Taking $1 - \alpha_1 = 0.9$  as the reference value, we can see that, due to temporal stability, in this case there is not a significant difference between the corresponding threshold values for the range of horizon times considered. Hence, the analysis will be performed under Method I adopting the fixed threshold $u_{1-\alpha_1} = u_{0.9}=2.9614$. 

\begin{figure}[ht!]
    \centering
    \includegraphics[width=0.33\textwidth]{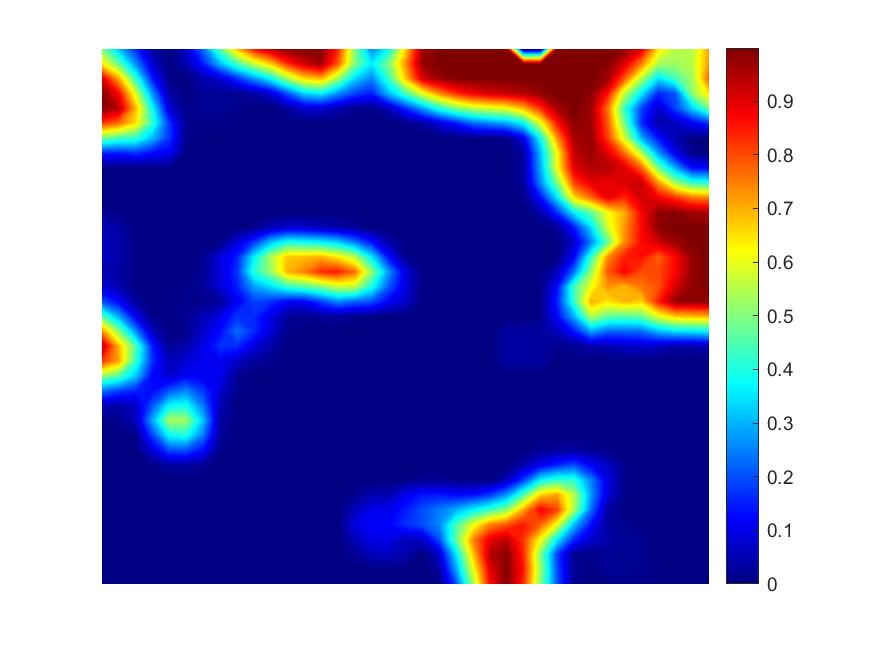}%
    \hfill
    \includegraphics[width=0.33\textwidth]{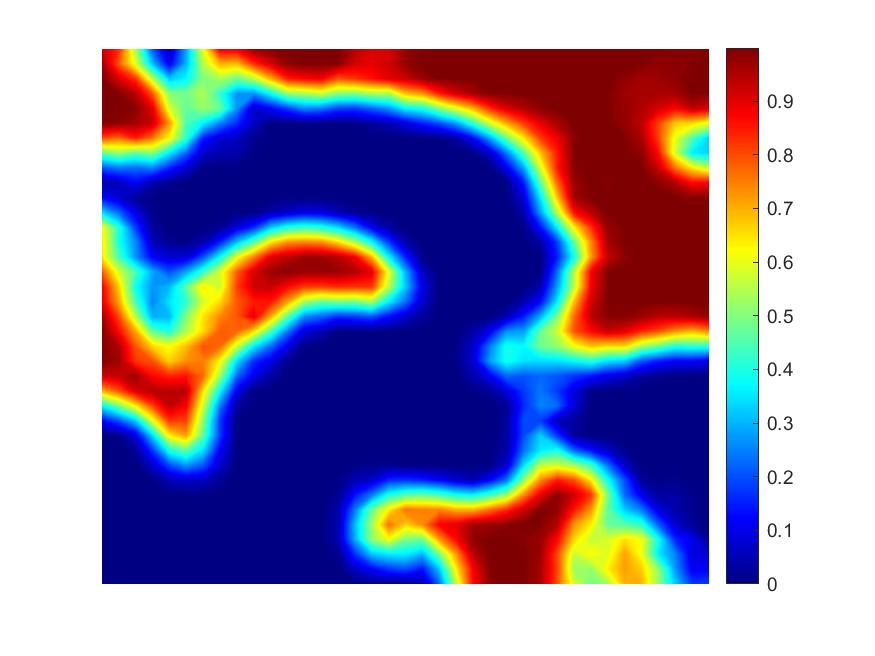}
  \hfill
    \includegraphics[width=0.33\textwidth]{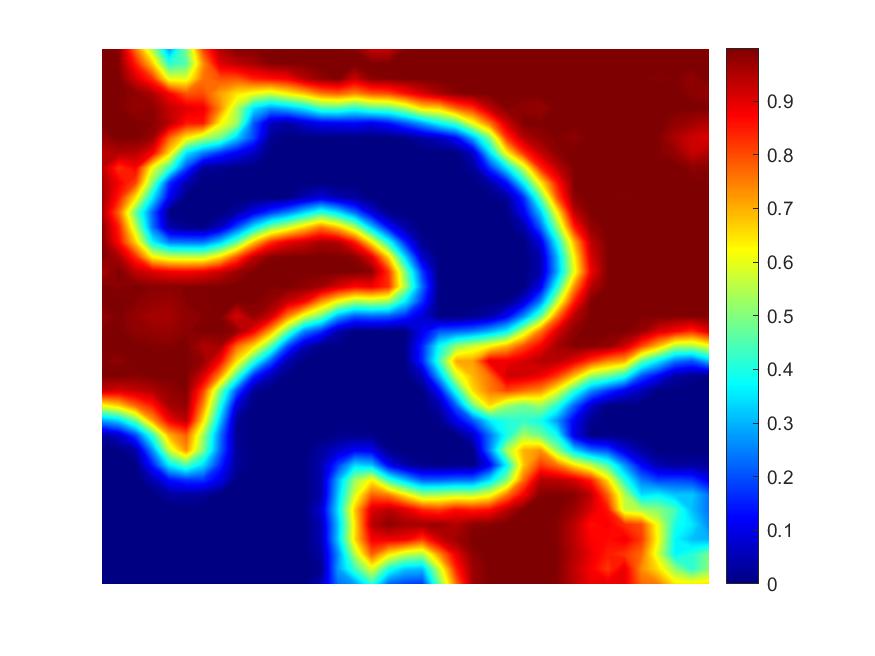}\\
    \includegraphics[width=0.33\textwidth]{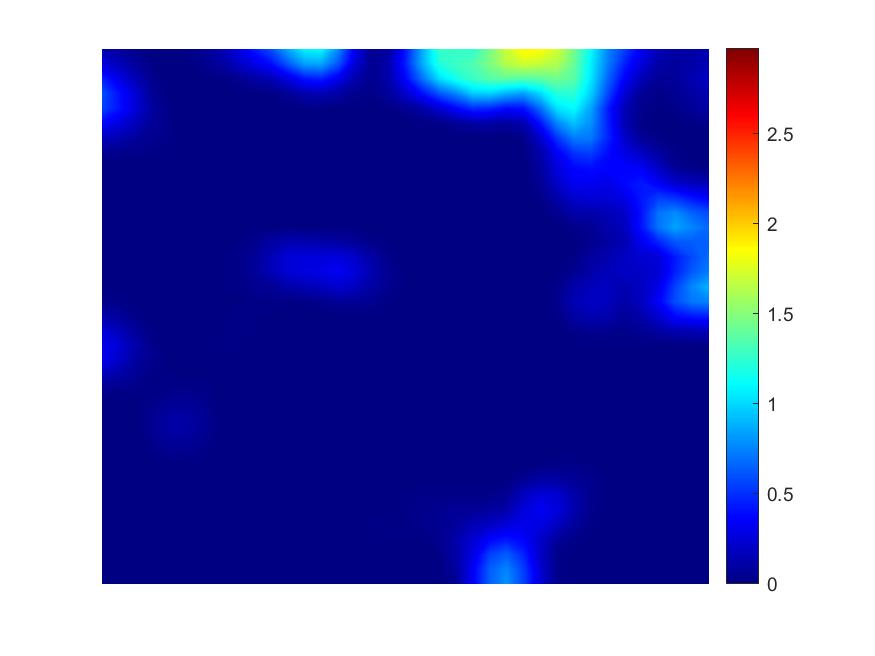}%
    \hfill
    \includegraphics[width=0.33\textwidth]{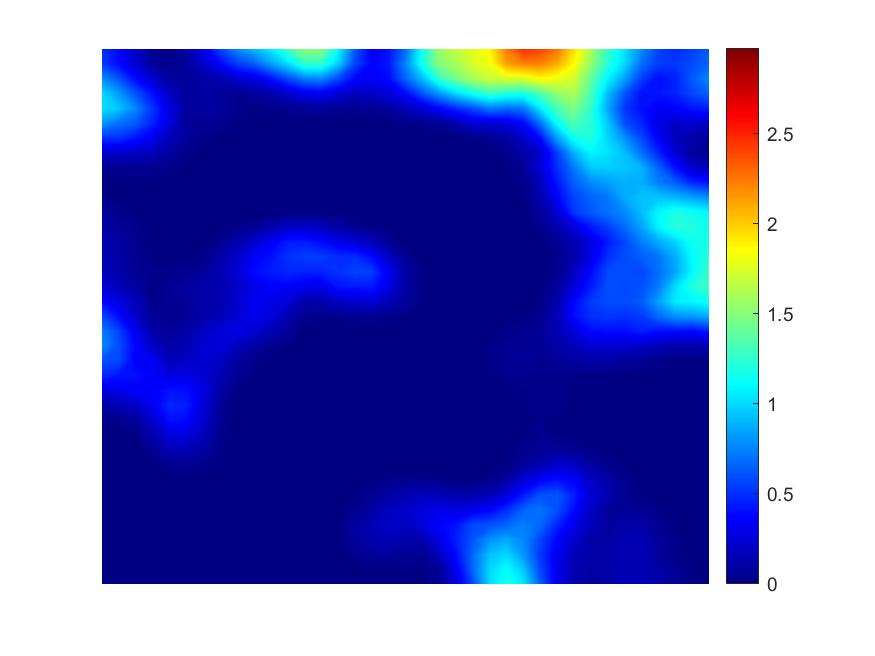}
  \hfill
    \includegraphics[width=0.33\textwidth]{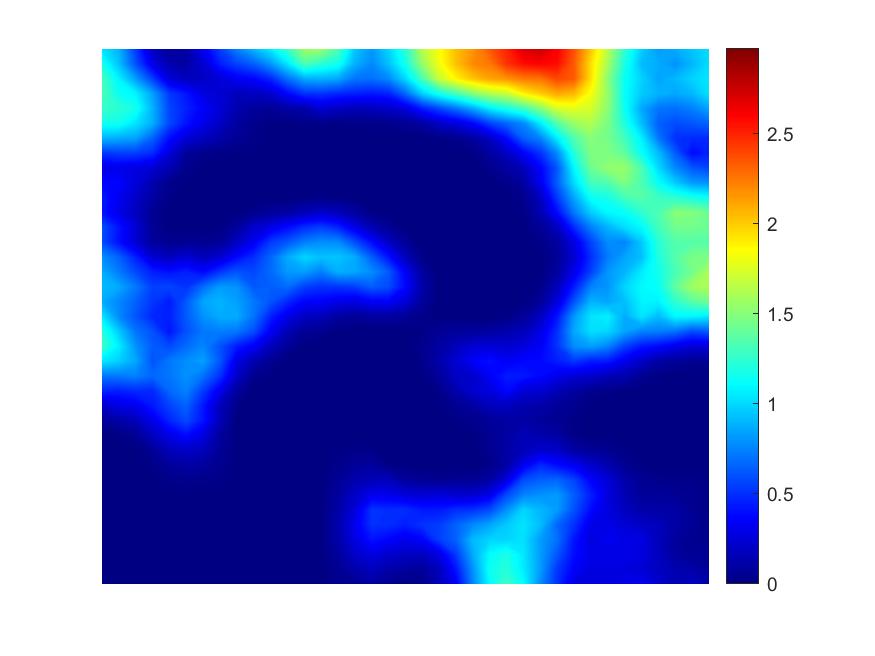}\\
    \includegraphics[width=0.33\textwidth]{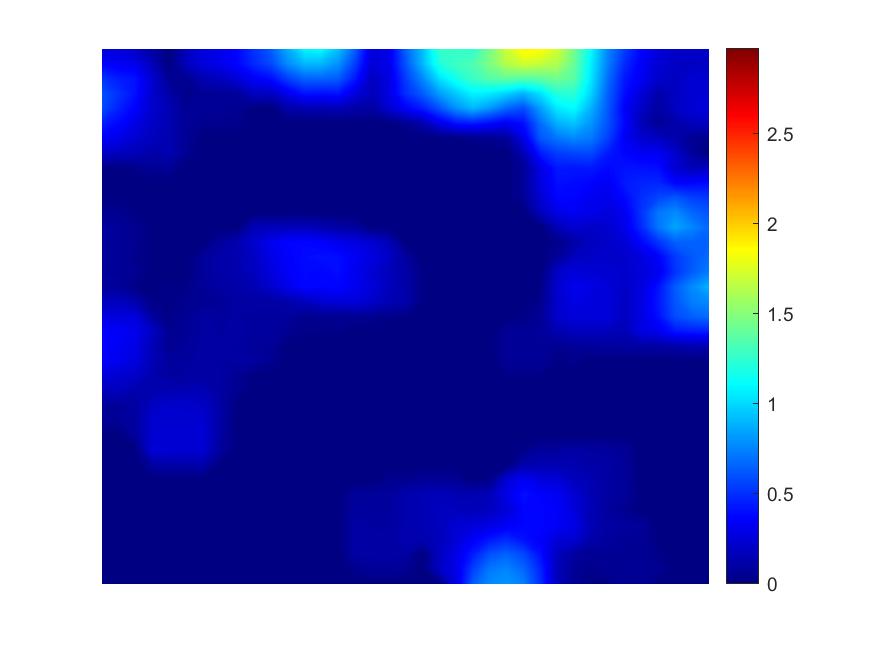}%
    \hfill
    \includegraphics[width=0.33\textwidth]{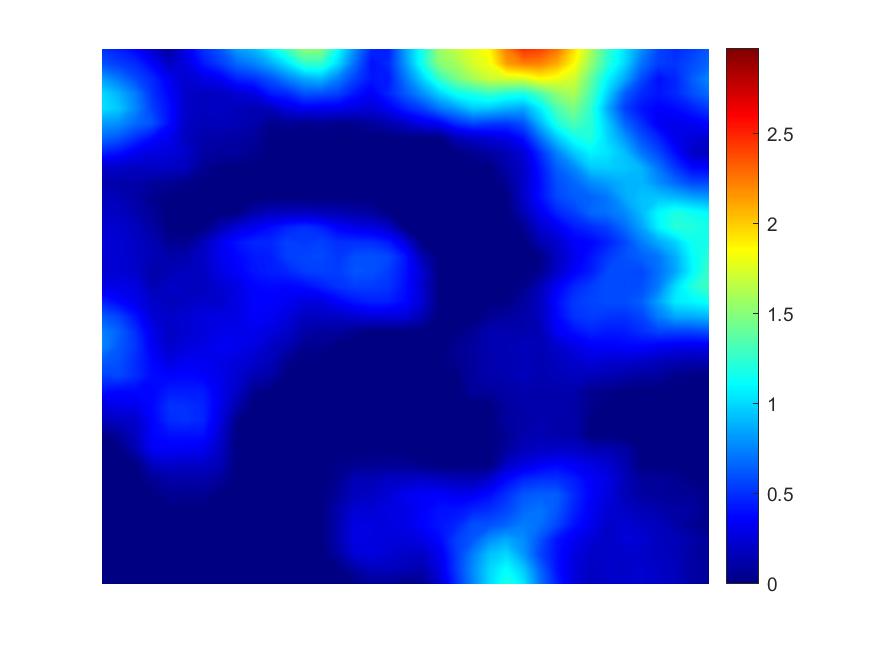}
  \hfill
    \includegraphics[width=0.33\textwidth]{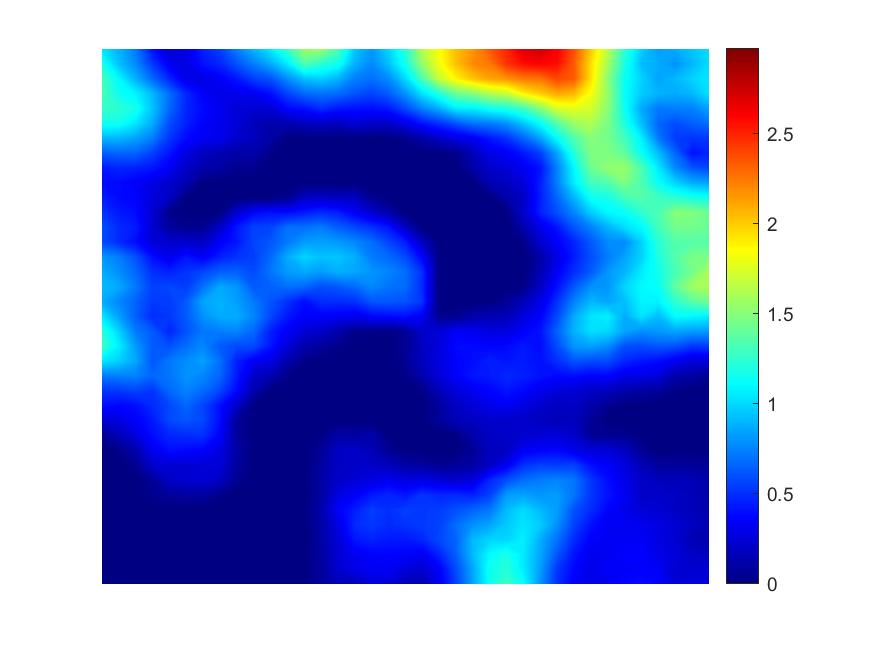}\\
  \caption{Realizations of the RCRFs using $VaR_{0.95}$ for the indicators: $\lambda^D$ (top), $V^D$ (middle), and $V^{A_u}$ (bottom). Columns correspond to horizon times $t=1,5,10$ (left to right)}
    \label{fig:VaR}
\end{figure}

\begin{figure}[ht!]
    \centering
    \includegraphics[width=0.33\textwidth]{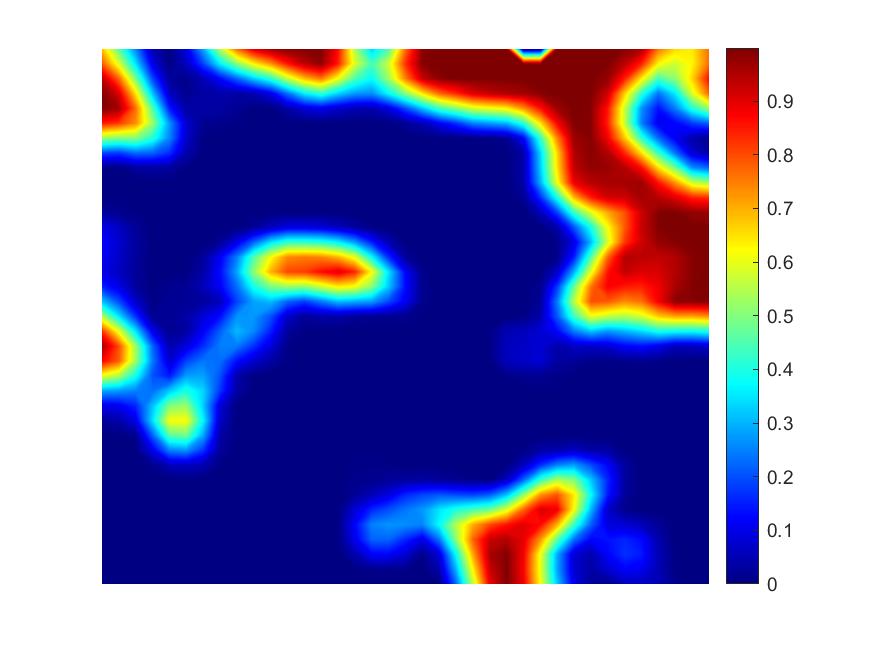}%
    \hfill
    \includegraphics[width=0.33\textwidth]{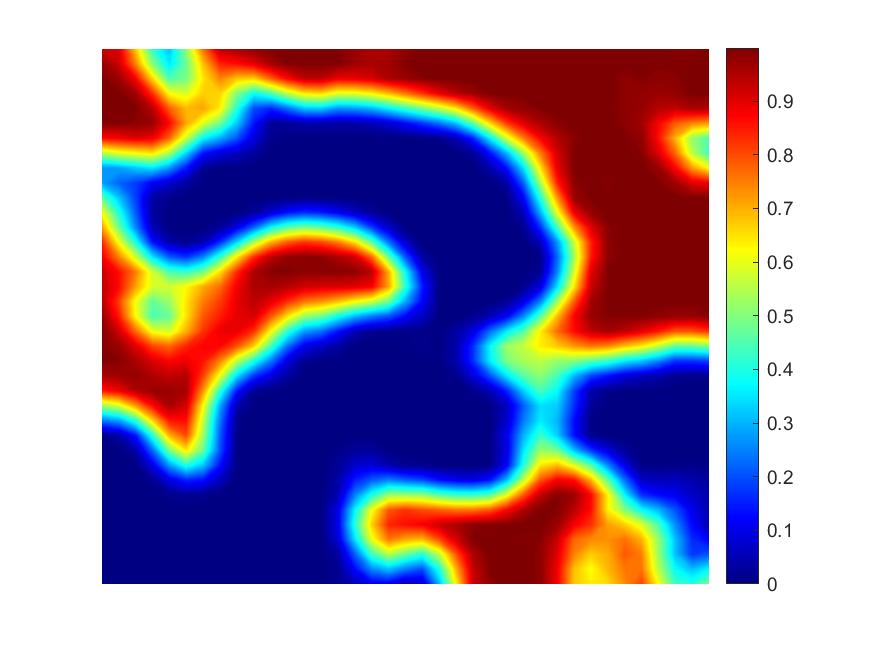}
  \hfill
    \includegraphics[width=0.33\textwidth]{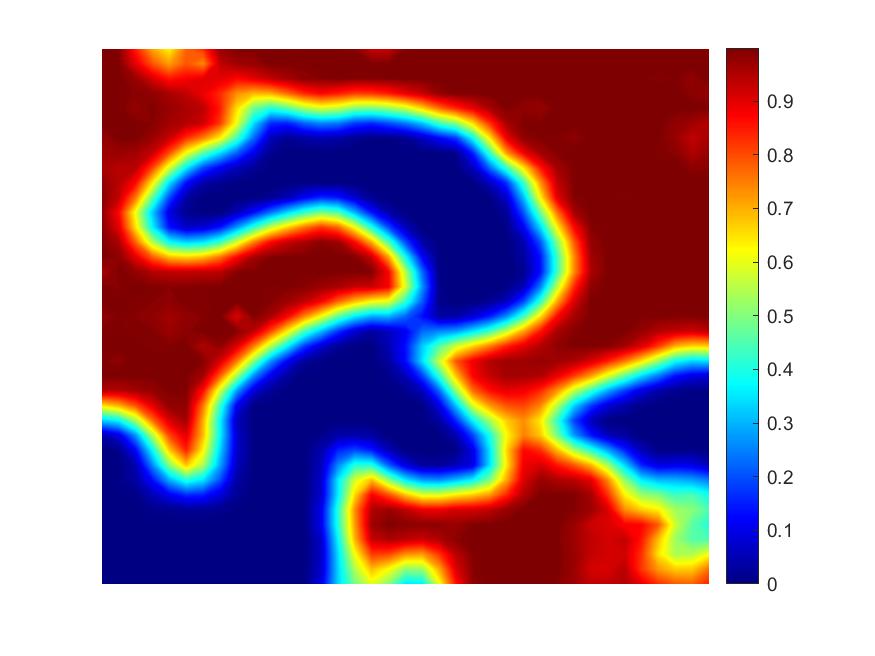}\\
    \includegraphics[width=0.33\textwidth]{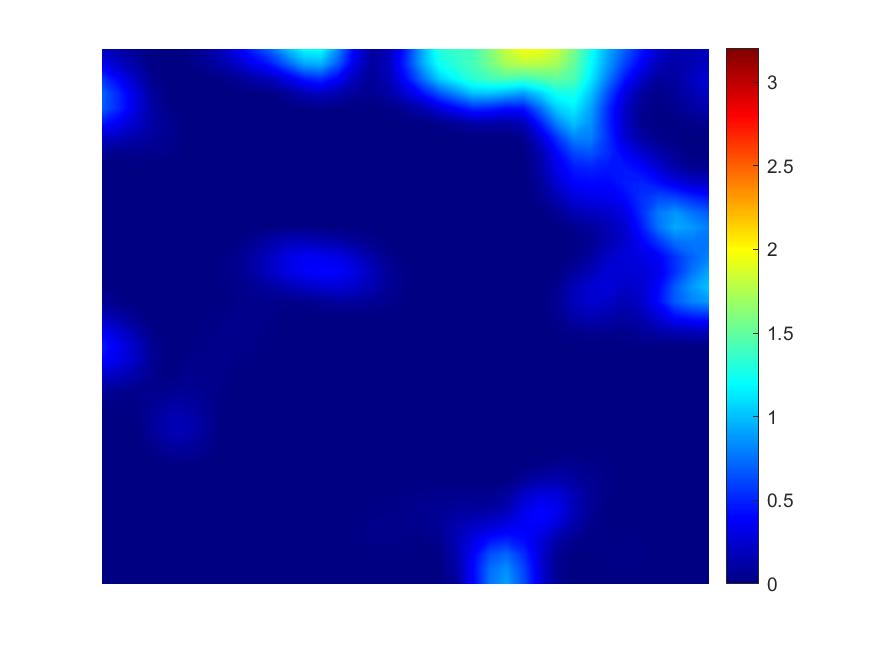}%
    \hfill
    \includegraphics[width=0.33\textwidth]{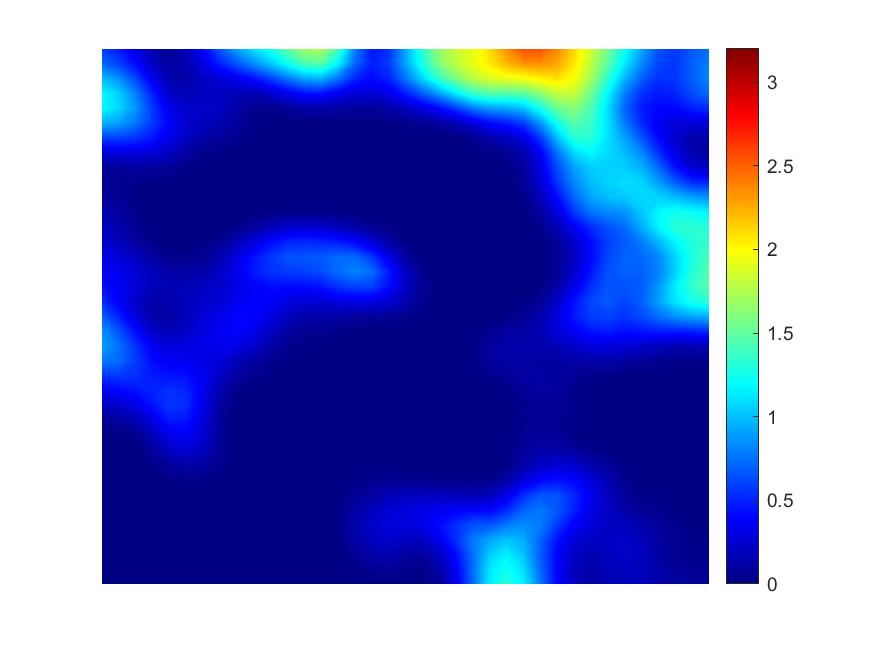}
  \hfill
    \includegraphics[width=0.33\textwidth]{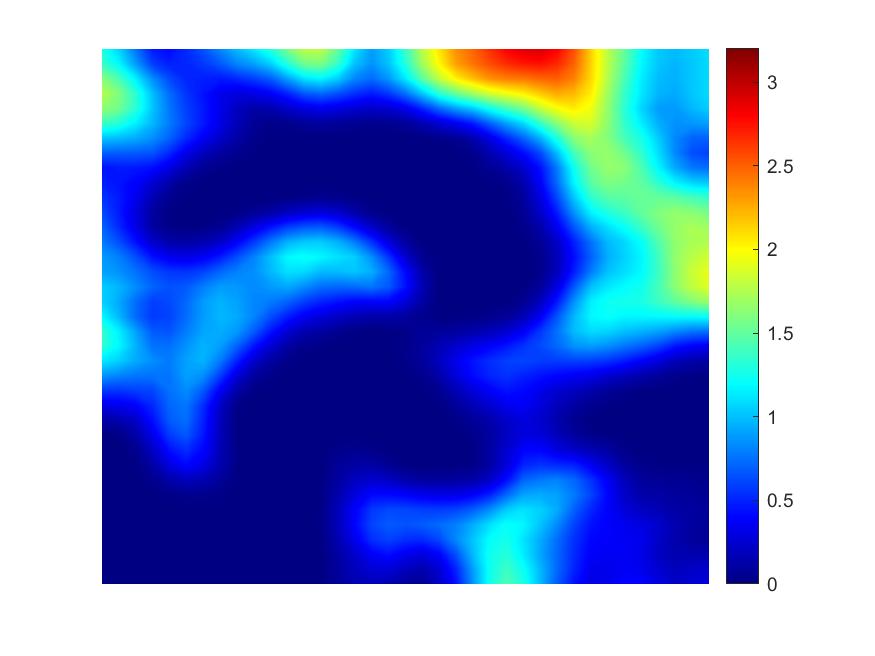}\\
    \includegraphics[width=0.33\textwidth]{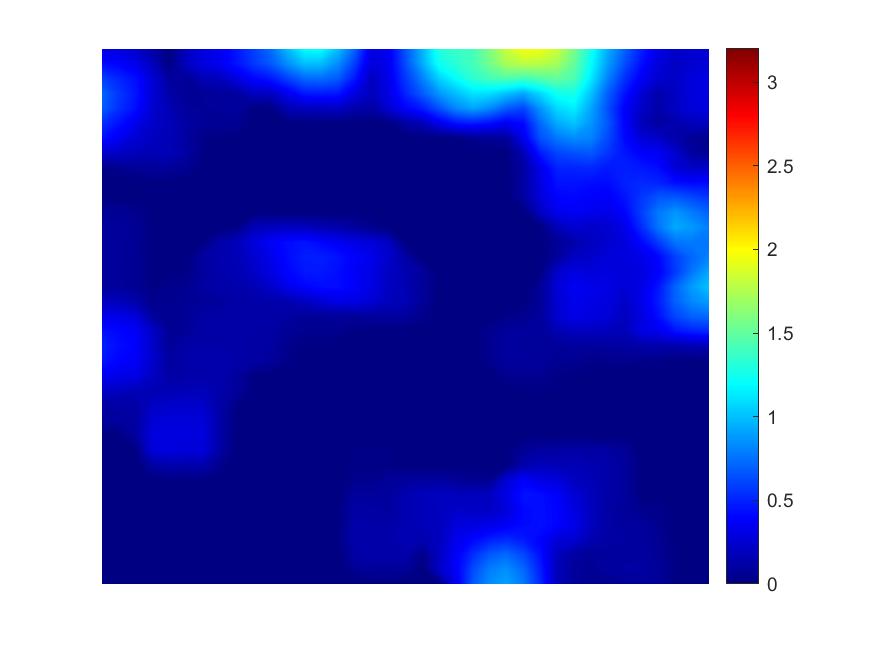}%
    \hfill
    \includegraphics[width=0.33\textwidth]{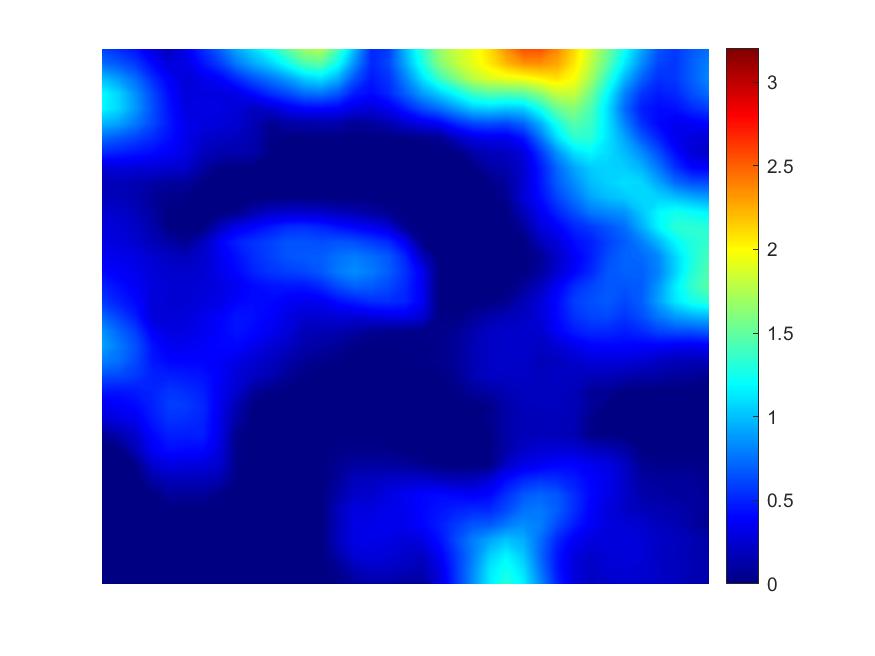}
  \hfill
    \includegraphics[width=0.33\textwidth]{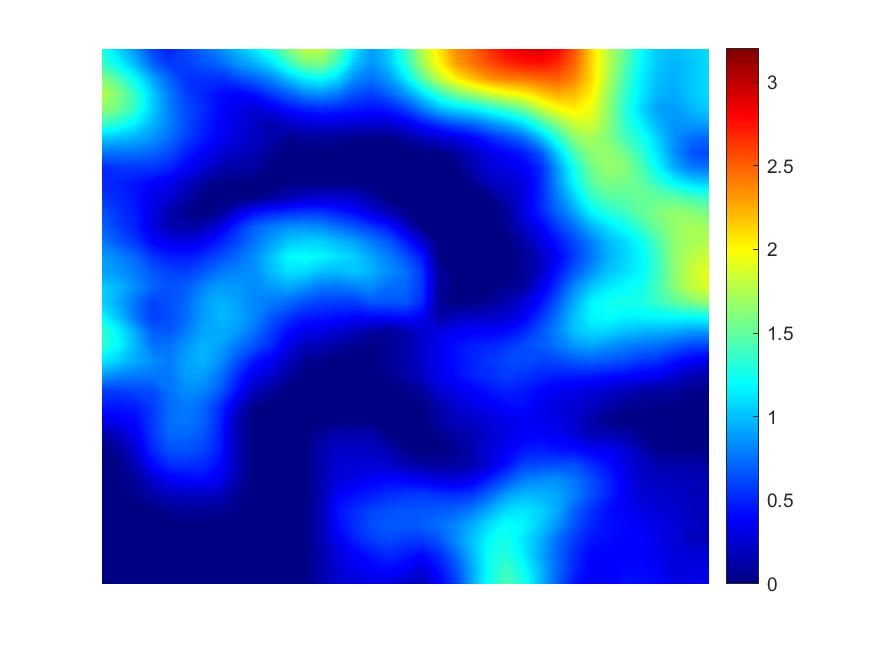}\\
  \caption{Realizations of the RCRFs using $AVaR_{0.95}$ for the indicators: $\lambda^D$ (top), $V^D$ (middle), and $V^{A_u}$ (bottom). Columns correspond to horizon times $t=1,5,10$ (left to right)}
    \label{fig:AVaR}
\end{figure}

In Figs. \ref{fig:VaR} and \ref{fig:AVaR}, the realizations of the RCRFs corresponding to the indicators $\lambda^D$, $V^D$ and $V^{A_u}$ (from top to bottom), respectively for risk measures $VaR_{0.95}$ and $AVaR_{0.95}$ (i.e., with $\alpha_2=0.05$), are displayed for horizon times $t=1,5,10$ (from left to right). As expected, the levels of risk increase with $t$, and AVaR is always locally larger than VaR for each selected indicator and horizon time. The dynamic properties of the spatial deformation $\Phi$ are clearly reflected in the evolution of the conditional risk fields realizations. In particular, a rotational movement of the risk structures is observed, with a progressive displacement towards the center of the domain as $t$ increases. Regarding the choice of a specific indicator, the area-based ($\lambda^D$) RCRF produces spatially more extended risk patterns, while  the volume-based ($V^D$ and $V^{A_u}$) RCRFs identify more localized and precise regions of higher risk intensity by accounting for the magnitude of exceedance.    

\begin{figure}[ht!]
    \centering
    \includegraphics[width=0.33\textwidth]{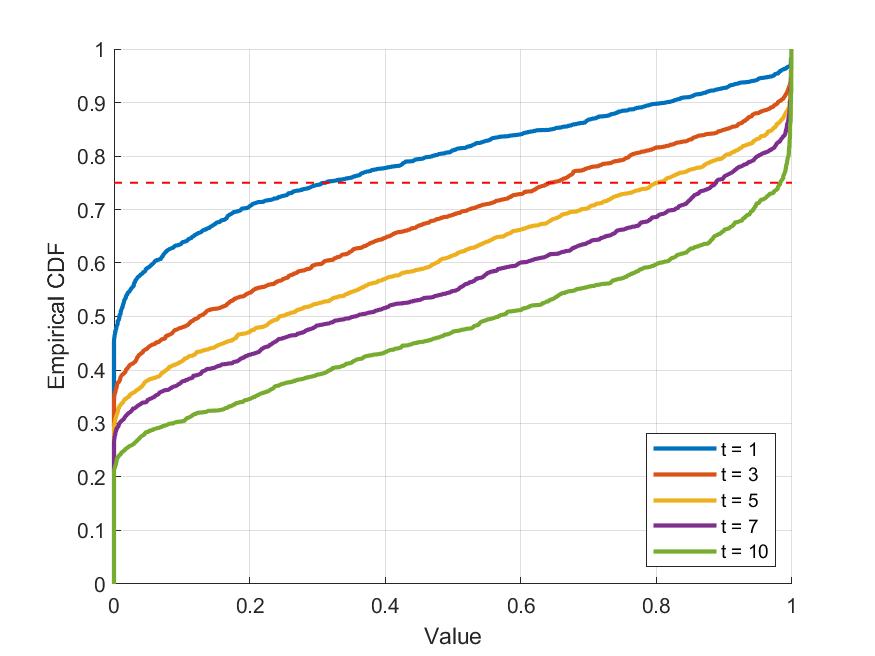}%
    \hfill
    \includegraphics[width=0.33\textwidth]{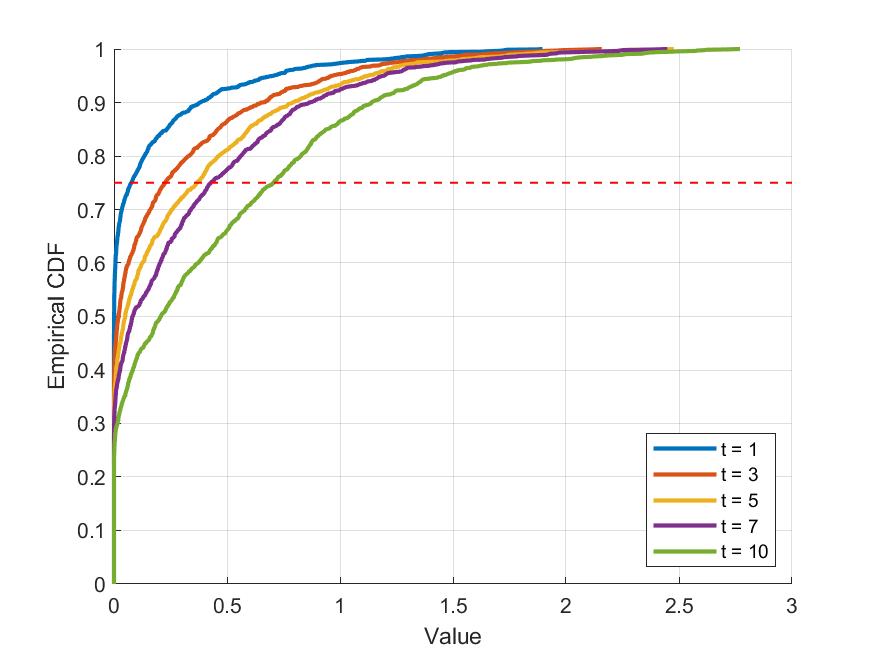}
  \hfill
    \includegraphics[width=0.33\textwidth]{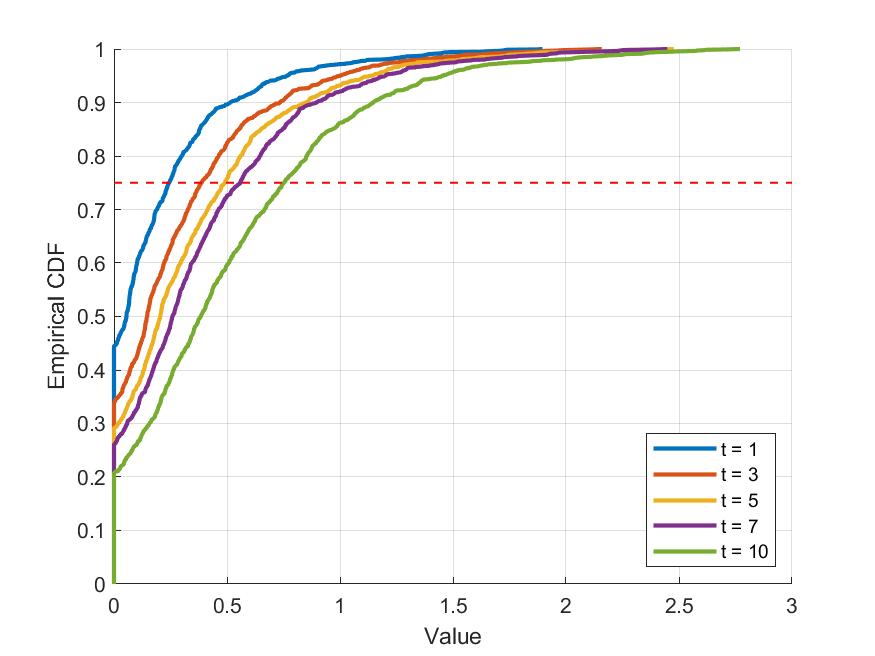}\\
    \includegraphics[width=0.33\textwidth]{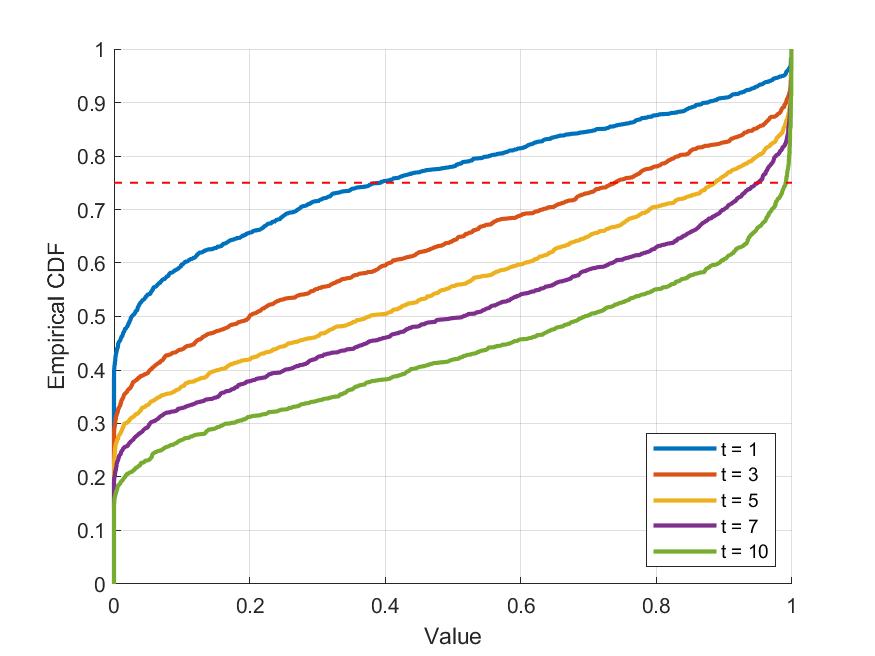}%
    \hfill
    \includegraphics[width=0.33\textwidth]{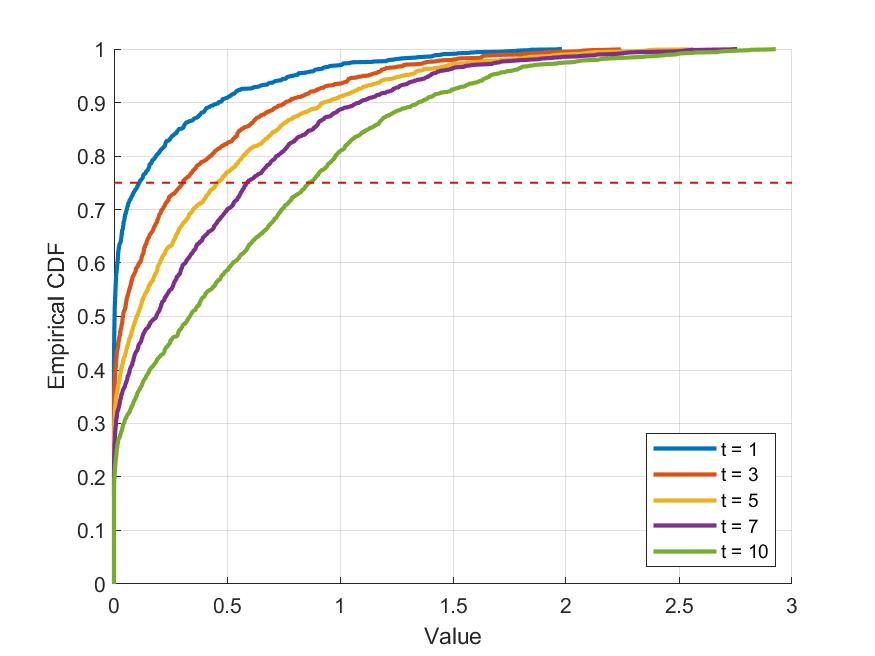}
  \hfill
    \includegraphics[width=0.33\textwidth]{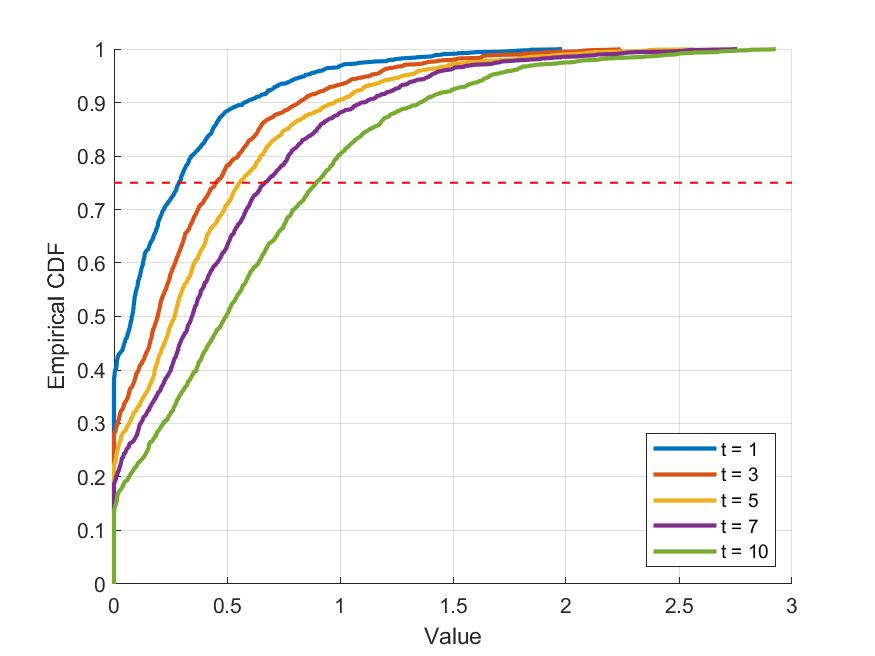}\\
   
   \caption{Empirical CDFs for $VaR_{0.95}$ (top row) and $AVaR_{0.95}$ (bottom row) for the indicators $\lambda^D$, $V^D$, and $V^{A_u}$ (left to right)}

    \label{fig:ECDF_u3}
\end{figure}

The previously mentioned effects are more evident in the empirical distributions represented in Fig. \ref{fig:ECDF_u3}, where each curve, for a given horizon time, is obtained from the values of the risk measure calculated on the selected indicator in the subdomains analyzed under the window design adopted. 
A comparison between the two rows of plots confirms that the AVaR distributions (bottom row) are consistently shifted toward higher values relative to the VaR ones (top row). Similarly, as $t$ increases, the curves exhibit greater dispersion, quantifying how the risk distribution becomes more heterogeneous across the domain. The distinction between indicators is also clearly reflected in the shape of the corresponding curves represented in the three columns of plots: the more extended risk patterns of $\lambda^D$ (left column) correspond to a gradual spread in its distribution, whereas the localized high-intensity regions identified by $V^D$ (central column) and $V^{A_u}$ (right column) are characterized by a more pronounced concentration of values in the upper tail, with the latter displaying particularly significant differences at the origin.      

\begin{figure}[htbp]
    \centering
    \includegraphics[width=0.33\textwidth]{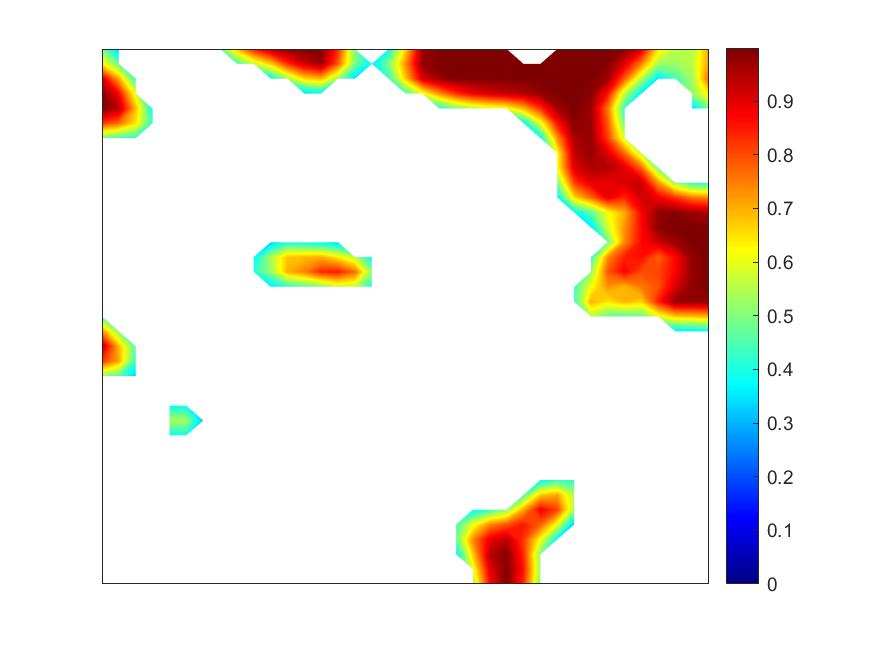}%
    \hfill
    \includegraphics[width=0.33\textwidth]{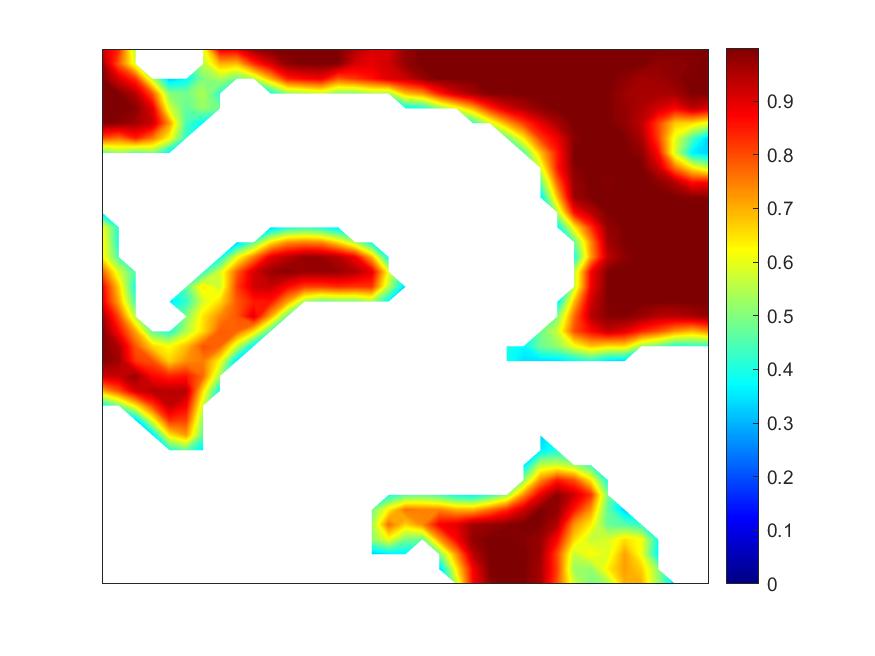}
  \hfill
    \includegraphics[width=0.33\textwidth]{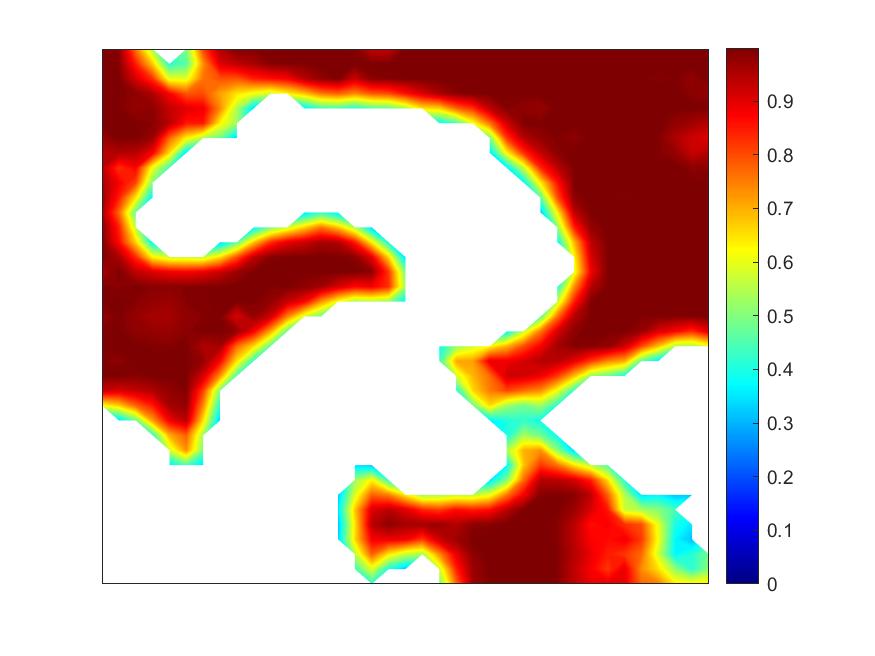}\\
     \includegraphics[width=0.33\textwidth]{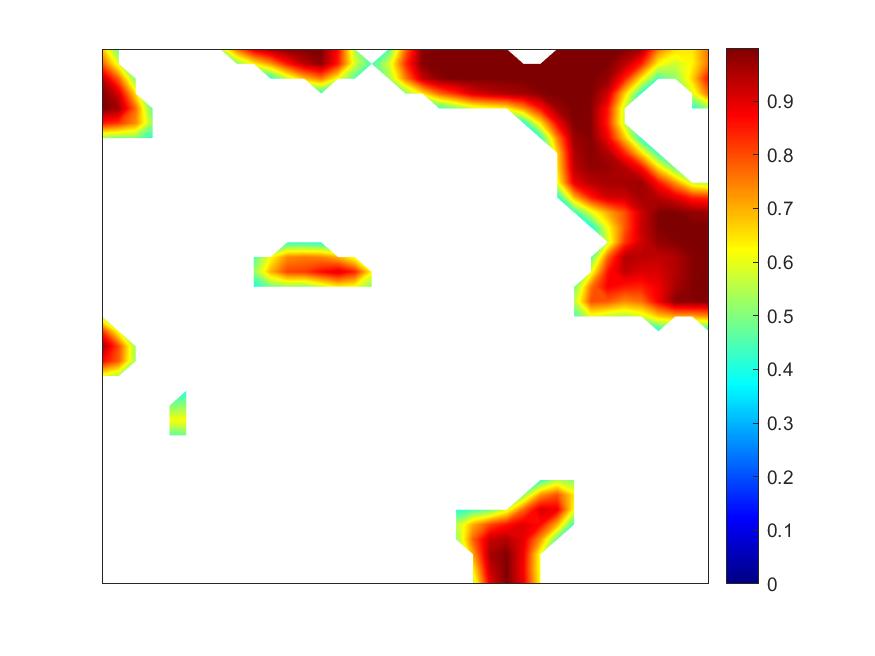}%
    \hfill
    \includegraphics[width=0.33\textwidth]{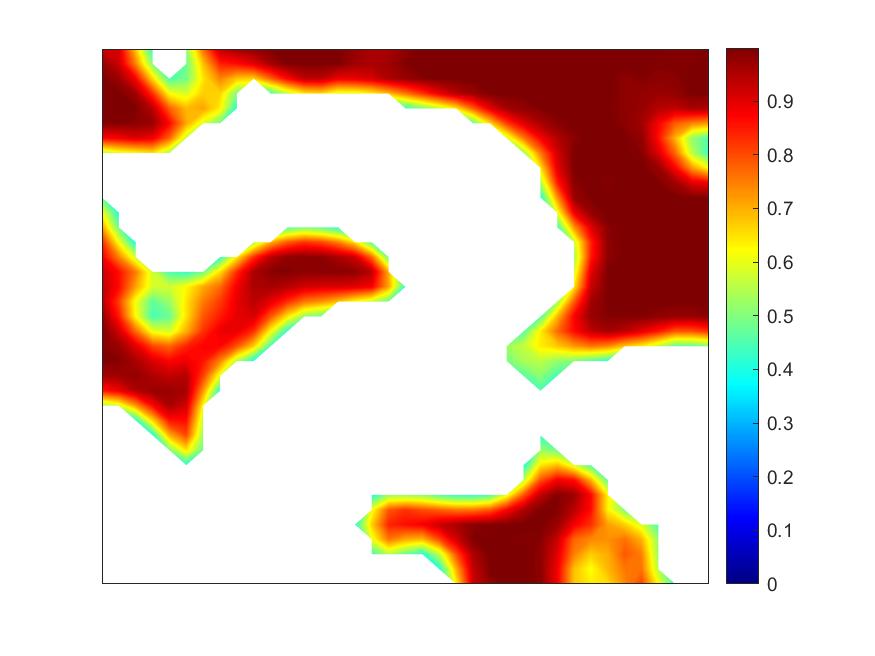}
  \hfill
    \includegraphics[width=0.33\textwidth]{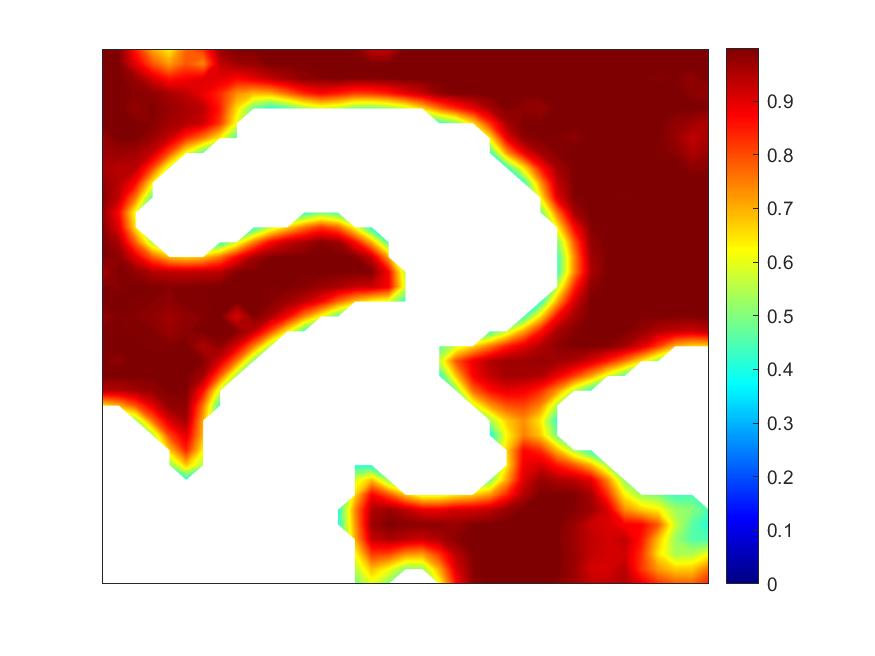}
\caption{Hotspots for $VaR_{0.95}(\lambda^D)$ (top row) and $AVaR_{0.95}(\lambda^D)$ (bottom row) at horizons $t=1,5,10$ (left to right), derived from the realizations shown in the top rows of Figs.~\ref{fig:VaR} and \ref{fig:AVaR}, respectively, using a fixed threshold $v_{1-\alpha}=v_{0.75}$ (Method I in Section~\ref{subsec:threshold})}
    \label{fig:Hotspots_MI}
    \end{figure}

\begin{figure}[htbp]   
    \includegraphics[width=0.33\textwidth]{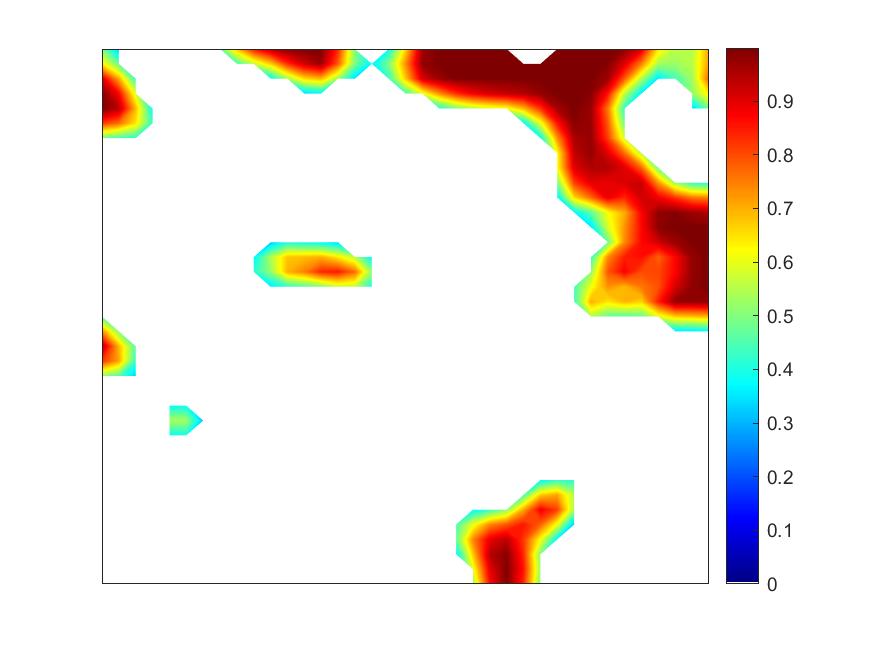}%
    \hfill
    \includegraphics[width=0.33\textwidth]{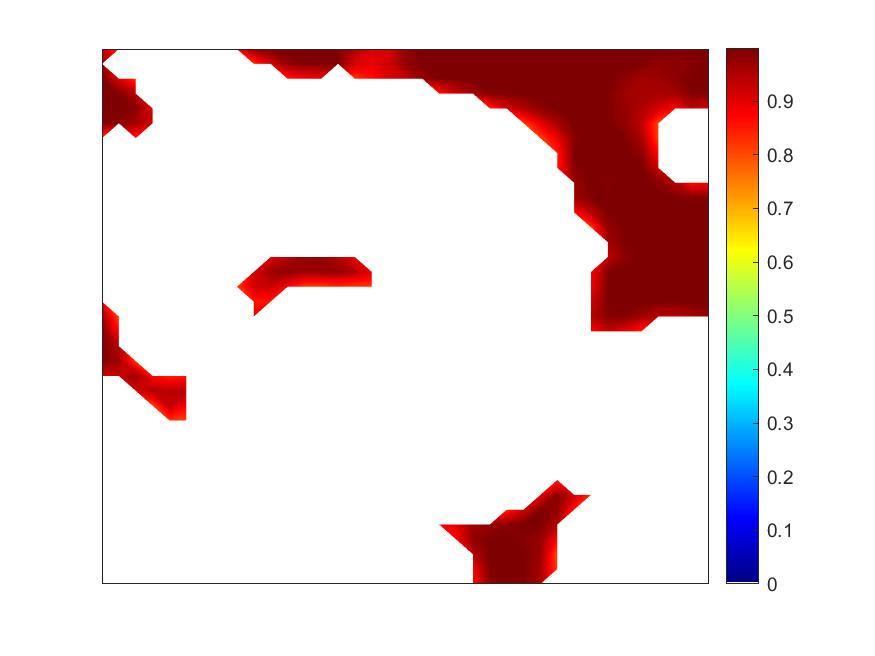}
  \hfill
    \includegraphics[width=0.33\textwidth]{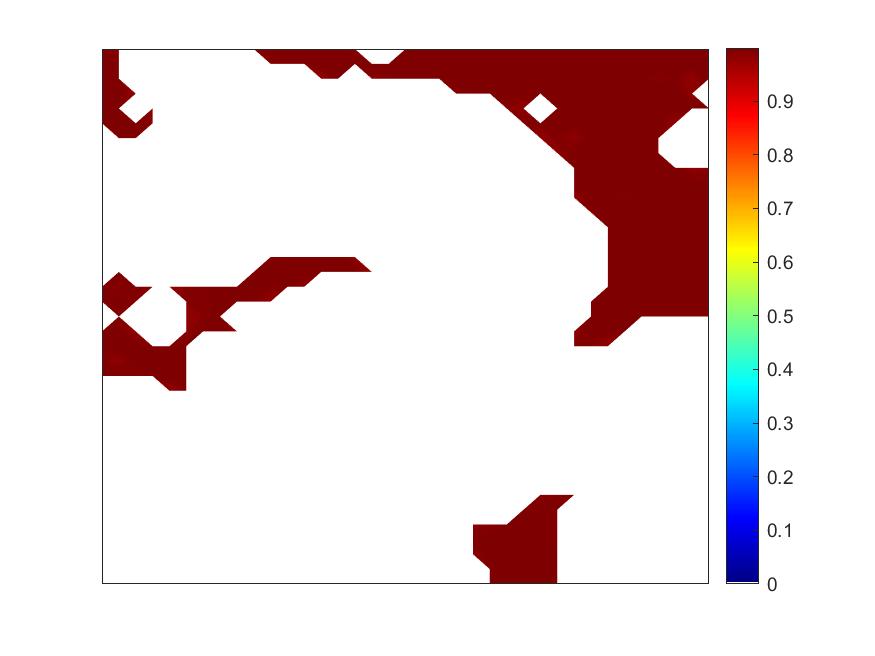}\\
      \includegraphics[width=0.33\textwidth]{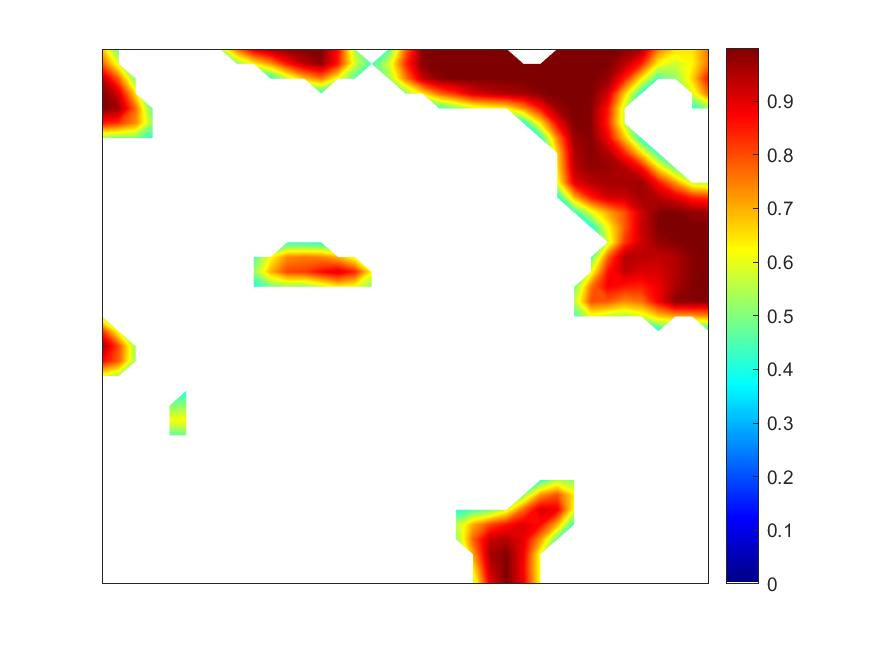}%
    \hfill
    \includegraphics[width=0.33\textwidth]{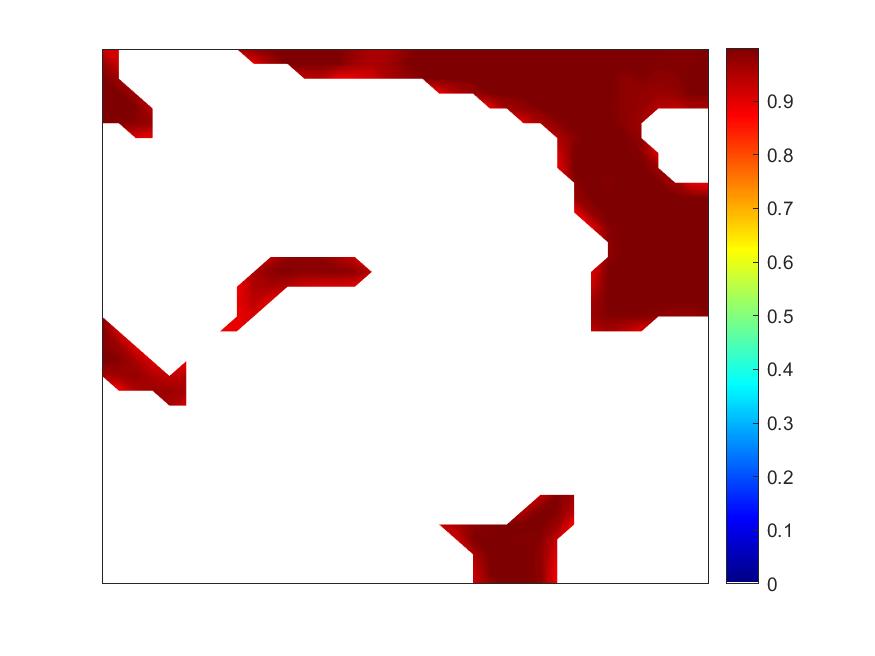}
  \hfill
    \includegraphics[width=0.33\textwidth]{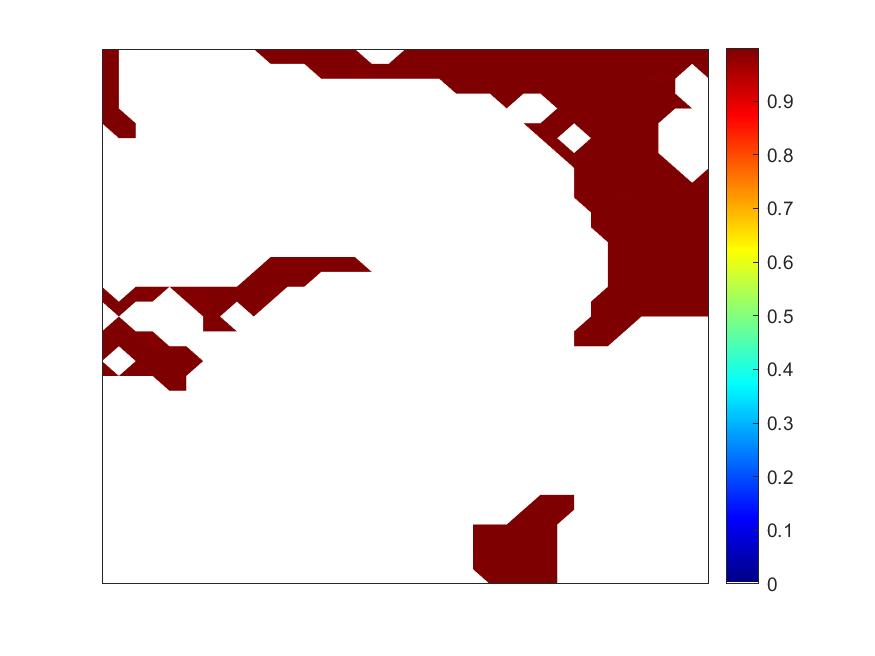}\\
  \caption{Hotspots for $VaR_{0.95}(\lambda^D)$ (top row) and $AVaR_{0.95}(\lambda^D)$ (bottom row) at horizons $t=1,5,10$ (left to right), derived from the realizations shown in the top rows of Fig.~\ref{fig:VaR} and \ref{fig:AVaR}, respectively, using a time-varying threshold $v_{t,1-\alpha}=v_{t,0.75}$ (Method II in Section~\ref{subsec:threshold})}
    \label{fig:Hotspots_MII}
\end{figure}

These curves are used to determine the thresholds defining hotspots, taken as their $0.75$-quantile according to the critical level considered. Under Method I, the threshold $v=v_{0.75}$ is based on the curve corresponding to $t=1$ and remains fixed for all the subsequent horizon times; Fig. \ref{fig:Hotspots_MI} displays some example hotspot maps obtained for VaR and AVaR of relative exceedance area at horizon times $t=1, 5, 10$.  With Method II, a threshold $v_t=v_{t,0.75}$ is calculated for each horizon time based on the corresponding empirical distribution; see related results in Fig. \ref{fig:Hotspots_MII}. 

As a result of these different definitions, the identified hotspots exhibit distinct temporal evolutions. In Method I, the global hotspots area increases significantly with $t$, as the fixed threshold $v$ (defined at $t=1$) is progressively exceeded due to the overall growth of the risk levels. This leads to the identification of increasingly larger regions where the risk surpasses the initial reference level. On the other hand, in Method II, the identified regions are more stable in size. Since the threshold $v_t$ is updated for each $t$, this method  highlights only those subregions  representing the highest risk relative to the current state of the domain. Thus, while Method I captures the absolute expansion of risk, Method II focuses on identifying the most critical locations at each specific horizon time, providing a normalized perspective of the hotspot dynamics.

\section{Conclusions}\label{sec:conclusions}

We introduce a general methodological approach for assessment of hotspot dynamics, from a risk-oriented perspective focused on structural aspects of threshold exceedances. More precisely, risk maps are built in reference to critical values of target risk measures, evaluated at subregional scales on selected indicators in terms of random geometrical characteristics of threshold exceedance sets. In practice, this methodology enables the generation of multiple output products, considering different choices for the input elements involved as tuning parameters, with specific inherent interpretations, thus having a complementary value for multitarget prospective risk evaluation and decision making. Technically, the procedure proposed is founded on the empirical approximation of the indicators probability distributions by model-based simulation conditional on available observations, leading to the formal definition of \emph{risk conditional random fields} (RCRFs). The realizations of the latter then provide \emph{risk-related hotspot} representations. Hence, hotspots in this general setup are defined by exceedances over critical levels of risk measures instead of the direct values of the physical magnitude of interest.                     

In our practical illustration of the methodology, we adopt a spatiotemporal model involving deformation of the spatial domain over time, with the aim of showing its dynamical effect on the risk-related hotspot maps at subsequent prediction horizons, under several specifications of
the RCRF input elements. 

Within this new framework, different directions can be identified for further research, among which we highlight the following. From a theoretical perspective, an important research line concerns the asymptotic behavior of RCRFs and their associated hotspots as the various critical levels involved tend towards extremal values, which would be affected by structural aspects of the reference spatiotemporal model, such as the presence of heavy tails. Possible methodological extensions, requiring developments on fundamental aspects, include the consideration of two-point and higher-order indicators, associated with underlying bivariate and multivariate compound cumulative distribution functions, as well as risk measures evaluated on multidimensional loss functions constructed from several indicators or multidimensional random fields; in this context, copulas provide a suitable approach for describing the joint dependence structure (among other references, see \citeauthor{Salvadori2007} \citeyear{Salvadori2007};
\citeauthor{Ruschendor2013} \citeyear{Ruschendor2013}). Exceedances can be generalized under functional and random (e.g., determined by covariate effects) threshold specifications. Finally, as practical advice for real data applications, a systematic sensitivity analysis with respect to the input elements of the RCRF would be especially interesting; key aspects include the scales
determined by the sliding-window design and the number of conditional simulations depending on the characteristics of the underlying model.

\bmhead{Acknowledgements}
This work was partially supported by  grant PGC2018-098860-B-I00 funded  by MICIU / AEI/10.13039/501100011033 and by ERDF A way of making Europe, grants  PID2021-128077NB-I00 and PID2025-171143NB-I00 funded  by MICIU / AEI/10.13039/501100011033 and by ERDF, EU, and grant CEX2020-001105-M funded by MICIU / AEI/10.13039/501100011033.

\bibliography{sn-bibliography}
\end{document}